\def\makeSM{1}
\documentclass[
bibnotes,
amsmath,amssymb,
pra,						     
twocolumn,                   
letterpaper,
twoside,
frontmatterverbose,
floatfix,
notitlepage,
]{revtex4-1}

\usepackage{amsmath}
\usepackage{mathtools}
\usepackage{graphicx}
\usepackage{dcolumn}
\usepackage{bm}
\usepackage{hyperref}
\usepackage{multirow}
\usepackage{array}
\usepackage{booktabs}
\usepackage{upgreek}
\usepackage{epsfig}
\usepackage{mathrsfs}
\usepackage{amssymb}
\usepackage{amsbsy}
\usepackage{color}
\usepackage{cancel}
\usepackage{pifont}
\usepackage{marginnote}
\usepackage{float}
\usepackage{tikz}
\usepackage{ctable}

\usetikzlibrary{arrows}
\tikzstyle{block}=[draw opacity=0.7,line width=1.4cm]
\usetikzlibrary{positioning}
\usepackage[caption=false]{subfig}
\usepackage{setspace}

\newcommand{\bcen}{\begin{center}}
\newcommand{\ecen}{\end{center}}
\newcommand{\btab}{\begin{tabular}}
\newcommand{\etab}{\end{tabular}}
\newcommand{\bdes}{\begin{description}}
\newcommand{\edes}{\end{description}}

\newcommand{\beq}{\begin{equation}}
\newcommand{\eeq}{\end{equation}}
\newcommand{\bea}{\begin{eqnarray}}
\newcommand{\eea}{\end{eqnarray}}

\newcommand{\bary}{\begin{array}}
\newcommand{\eary}{\end{array}}
\newcommand{\benum}{\begin{enumerate}}
\newcommand{\eenum}{\end{enumerate}}
\newcommand{\bitem}{\begin{itemize}}
\newcommand{\eitem}{\end{itemize}}

\newcommand{\D}[1]{\mbox{d}{#1}}

\newcommand{\paratitle}[1]{\noindent {\bf #1:}}

\newcommand{\eqn}[1] {eqn.~(\ref{#1})}

\newcommand{\fig}[1]{fig.~\ref{#1}}

\newcommand \SMcite[1]{SM \ref{#1}}

\makeatletter

\newcommand{\Rmnum}[1]{\expandafter\@slowromancap\romannumeral #1@}
\makeatother

\newcommand{\mylabel}[1]{\label{#1}} 

\newcommand{\mycite}[1]{\cite{#1}}

\newcommand{\authorprl}[2]{\author{#1}\email{#2}}

\newcommand{\qc}{{q_c}}

\newcommand{\cH}{\mathscr{H}}

\newcommand{\ci}{\mathfrak{i}}

\newcommand{\dtau}{\textup{d}\tau}
\newcommand{\de}{\textup{d}\varepsilon}
\newcommand{\iwn}{\ci\omega_n}

\newcommand{\wn}{\omega_n}
\newcommand{\dt}{\textup{d}t}
\newcommand{\dw}{\textup{d}\omega}

\newcommand{\scdots}{{\tiny\cdots\normalsize}}
\newcommand{\cd}{c^\dagger}
\newcommand{\cb}{\bar{c}}
\newcommand{\ek}{\varepsilon_{b,\mathbf{k}}}
\newcommand{\vk}{\mathbf{k}}
\newcommand{\Gkwn}{G_b(\iwn,\mathbf{k})}

\newcommand{\Abkw}{\rho_b(\omega,\mathbf{k})}
\newcommand{\db}{\bar{d}}
\newcommand{\LHS}{\textup{LHS}}
\newcommand{\FOne}{F^{\textup{(1)}}}
\newcommand{\FTwo}{F^{\textup{(2)}}}
\newcommand{\KOO}{K^{\textup{(11)}}}
\newcommand{\KTT}{K^{\textup{(22)}}}
\newcommand{\KOT}{K^{\textup{(12)}}}
\newcommand{\KTO}{K^{\textup{(21)}}}
\newcommand{\Kintra}{\tilde{K}_{\textup{intra}}}
\newcommand{\Kinter}{\tilde{K}_{\textup{inter}}}
\newcommand{\Kzintra}{\tilde{K}_{\textup{intra}}^{(q=0)}}
\newcommand{\Kzinter}{\tilde{K}_{\textup{inter}}^{(q=0)}}
\newcommand{\MKzintra}{\mathbf{K}_{\textup{intra}}^{(q=0)}}
\newcommand{\MKzinter}{\mathbf{K}_{\textup{inter}}^{(q=0)}}
\newcommand{\lzintra}{\lambda_{\textup{L(intra)}}^{(\vtr{q}=\vtr{0})}}
\newcommand{\lzinter}{\lambda_{\textup{L(inter)}}^{(\vtr{q}=\vtr{0})}}
\newcommand{\lmax}{\lambda^{(M)}_\textup{L}}
\newcommand{\vtr}[1]{\mathbf{#1}}
\newcommand{\Tr}{\textup{Tr}}
\newcommand{\NFL}{\textup{NFL}}
\newcommand{\LFL}{\textup{LFL}}
\newcommand{\act}{\textup{S}}
\renewcommand{\Im}{\mathrm{Im}}
\renewcommand{\Re}{\mathrm{Re}}

\newcommand{\figwidth}{0.99\columnwidth}

\newcommand{\titlename}{Higher-dimensional SYK Non-Fermi Liquids at Lifshitz transitions}
\newcommand{\showfigures}

\begin{document}


\title{\titlename}
\authorprl{Arijit Haldar}{arijithaldar@iisc.ac.in}
\authorprl{Sumilan Banerjee}{sumilan@iisc.ac.in}
\authorprl{Vijay B. Shenoy}{shenoy@iisc.ac.in}
\affiliation{Centre for Condensed Matter Theory, Department of Physics, Indian Institute of Science, Bangalore 560 012, India}


\date{\today}


\begin{abstract} 
We address the key open problem of a higher dimensional generalization of the Sachdev-Ye-Kitaev (SYK) model. We construct a model on a lattice of SYK dots with non-random intersite hopping. The crucial feature of the resulting band dispersion is the presence of a Lifshitz point where two bands touch with a tunable powerlaw divergent density of states (DOS). For a certain regime of the powerlaw exponent, we obtain a new class of interaction-dominated non-Fermi liquid (NFL) states, which  exhibits exciting features such as a zero-temperature scaling symmetry, an emergent (approximate) time reparameterization invariance, a powerlaw entropy-temperature relationship, and a fermion dimension that depends continuously on the DOS exponent. Notably, we further demonstrate that these NFL states are fast scramblers with a Lyapunov exponent $\lambda_L\propto T$, although they do not saturate the upper bound of chaos, rendering them truly unique.    
\end{abstract}

\maketitle

Description of quantum many-body systems lacking qausiparticle excitations \mycite{SachdevBook} is a longstanding and  challenging problem at the forefront of physics research today. A solvable instance of such a system in 0-dimension is provided by the SYK model \mycite{Sachdev1993,KitaevKITP}, which has attracted a lot of attention recently due to its intriguing connections to quantum gravity in $\mathrm{AdS_2}$ \mycite{Sachdev2010,KitaevKITP,Sachdev2015,Polchinski2016,Maldacena2016} and intertwined questions of thermalization and information scrambling. The exciting possibility of generalizing this model to higher dimensions to address questions relating to transport without quasiparticles as well as to look for possible dual to higher-dimensional gravity, has lead to a number of interesting extensions \mycite{Gu2016,SJian2017b,CJian2017,Song2017,Berkooz2016,Davison2017,Murugan2017,SJian2017a} of SYK model.

The attempts to generalize SYK model to higher dimensions typically use a lattice of SYK dots, connected via either random interdot interaction and/or hopping \mycite{Gu2016,SJian2017b,CJian2017,Song2017,SJian2017a} or through uniform hopping leading to a translationally invariant system \mycite{Zhang2017}. These generalizations have lead to extraction of transport quantities and diagnostics of many-body quantum chaos, such as the butterfly velocity \mycite{Gu2016},  in strongly interacting lattice models, demonstrating their connection, e.g., with the phenomena like heavy-Fermi liquids \mycite{Song2017} and many-body localization \mycite{SJian2017b}. However, such extensions have met with only partial success towards a `canonical' higher-dimensional generalization. In particular, the ensuing low-energy behavior turns out to be either qualitatively similar
to that of the 0-dimensional SYK model at the leading order \mycite{Gu2016}, or results in a low-temperature phase where interaction becomes irrelevant \mycite{Song2017,Zhang2017} at low energies. A notable higher-dimensional extension of SYK model by Berkooz et al. \mycite{Berkooz2016} has succeeded in obtaining an interaction-dominated fixed point distinct from 0-dimensional model by allowing SYK interactions only to low momentum fermions by using a phenomenological `filter function' construct whose microscopic underpinnings are unclear. Such issues have left the higher-dimensional extension of the SYK model as an interesting open problem. Here we  propose and  extensively study a microscopically motivated, translationally invariant lattice model, exhibiting a new class of interaction-dominated non-Fermi liquid (NFL) fixed points, derived from the SYK model.

\ifdefined\showfigures
\begin{figure}[h!]
\includegraphics[width=\figwidth]{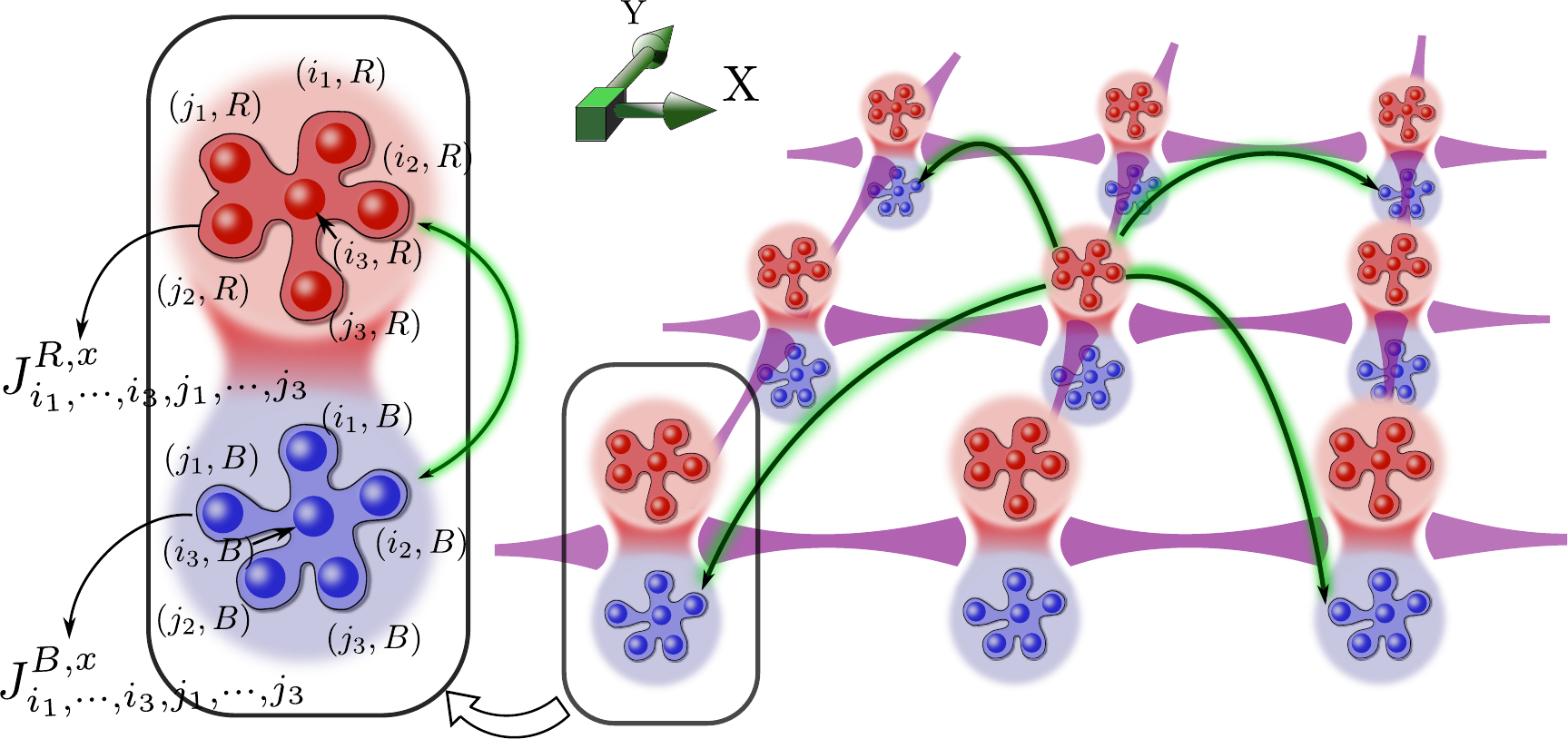}
\caption{{\bf Model:} The model consists of a $d$-dimensional hypercubic lattice whose unit cell comprises of SYK dots with $q$-body interactions of two colors(indexed R(red) and B(blue)), each with $N$-flavors of complex fermions. Translationally invariant hoppings(represented by green highlighted arrows) preserve the SYK flavor but flip the color index. Schematic shown here corresponds to $d=2$ and $q=3$.
}
\mylabel{fig:flagship_schem}
\end{figure}
\fi 

As illustrated in \fig{fig:flagship_schem}, our model comprises of SYK dots, with random intradot $q$-fermion
interactions \mycite{Maldacena2016},  at the sites of a $d$-dimensional lattice with $\emph{uniform}$ interdot hoppings. This leads to a dispersion with two particle-hole symmetric bands. The hoppings can be chosen such that the bands touch, as in a Lifshitz transition \mycite{Lifshitz1930}, with low-energy dispersion $\varepsilon_\pm(\mathbf{k})\propto \pm |\mathbf{k}|^p$ ($p\geq d$). This gives rise to a particle-hole symmetric density of states (DOS) with a powerlaw divergence, namely $g(\varepsilon)\sim|\varepsilon|^{-\gamma}$ with $0\leq \gamma=1-d/p < 1$. Such DOS singularity, e.g., may arise at topologically protected Fermi points with additional symmetries \mycite{Heikkila2010}.
This singularity in DOS is what ultimately enables the amalgamation
of lattice and SYK physics, thereby producing the family of new fixed
points mentioned earlier. The remarkable feature of this model is
that by tuning $\gamma$ a variety of fermionic phases can be realized. For values of
$\gamma$ less than a critical value $\gamma_{c}=(2q-3)/(2q-1)$, we
get a Fermi liquid like phase where interaction effects are only perturbative, while on the other hand, when $\gamma\approx1$ the system behaves similar to the parent SYK model. The most interesting physics occur for $\gamma_{c}<\gamma<1$, where the emergent NFL phases correspond to a new set of fixed points with a continuously tunable fermion dimension $\Delta=(1+\gamma)/2(1-\gamma+2\gamma q)$. As a result, the NFL phases have properties tunable via $\gamma$, and distinct from that of the parent SYK model. We find that unlike the pure SYK phase, the NFL phases have zero ground-state entropy ($S(T=0)$), with $S(T)$ varying as a powerlaw in $T$ with $\gamma$ dependent exponent. Moreover, the onset of quantum chaos in the model is governed by a Lyapunov exponent $\lambda_\mathrm{L}\propto T$, which, however, does not saturate the chaos bound of $\lambda_{L}=2\pi T$  \mycite{Maldacena2015,KitaevKITP}. As $\gamma\rightarrow1$, the residual zero-temperature entropy of the parent SYK is recovered and $\lambda_{L}\rightarrow 2\pi T$.

The truly higher-dimensional nature of the model is manifested in the non-trivial dynamical scaling exponent $z$ of the fermions. The perturbative fixed point ($\gamma<\gamma_c$) retains $z=p$ of the non-interacting model. Whereas, for $\gamma>\gamma_c$, interaction changes the dynamical exponent from $z=p$ to $z=p/(2(2q-1)\Delta-1)$, not unlike a proposed quantum gravity theory \mycite{Horava2009} which exploits Lifshitz points. Thus our work offers a solution to the much sought after higher-dimensional generalization
of the SYK model and provides a framework for addressing problems ranging
from the nature of transport and thermalization in systems without
quasiparticles to possible realizations of higher-dimensional duals
of quantum gravity models.

\paratitle{Model} Our lattice model (\fig{fig:flagship_schem} ) is described by the following Hamiltonian
\begin{align}\mylabel{eqn:Ham}
\cH =&-\sum\limits_{\mathbf{x},\mathbf{x'},\alpha,\alpha'}t_{\alpha\alpha'}(\mathbf{x}-\mathbf{x'})c^\dagger_{i\alpha\mathbf{x}}c_{i\alpha'\mathbf{x'}}
-\mu\sum\limits_{\mathbf{x},i\alpha}
c^\dagger_{i\alpha\mathbf{x}}c_{i\alpha\mathbf{x}}\notag\\
+&\sum\limits_{\mathbf{x},\alpha}\ \ \ \sum\limits_{\mathclap{\substack{i_1,\cdots,i_{q},\\j_1,\cdots,j_{q}}}}  J^{\alpha,\mathbf{x}}_{i_1,\cdots,i_{q};j_1,\cdots,j_{q}}c^\dagger_{i_{q}\alpha\mathbf{x}}\cdots c^\dagger_{i_1\alpha\mathbf{x} } c_{j_1\alpha \mathbf{x}} \cdots c_{j_q\alpha\mathbf{x}},
\end{align}
where $i,j=1,\cdots,N$ denote SYK flavor at the lattice point $\mathbf{x}$ in $d$ dimension and $\alpha,\alpha'=R$(red)$,B$(blue) index the two colors for the fermion operators $c,c^\dagger$. The intradot complex random all-to-all $q$-body SYK couplings $J^{\alpha,\mathbf{x}}_{i_1,\cdots,i_{q};j_1,\cdots,j_{q}}$ are completely local and scatter fermions having the same color index  (see \fig{fig:flagship_schem}). These amplitudes are identically distributed independent random variables with variance $J^{2}/qN^{2q-1}(q!)^2$, such that
$\langle J^{\alpha,\mathbf{x}}_{i_1,\scdots,i_{q};j_1,\scdots,j_{q}}J^{\alpha,\mathbf{x'}}_{i'_1,\scdots,i'_{\qc};j'_1,\scdots,j'_{q_c}}\rangle
\propto 
\delta_{\alpha,\alpha'}\delta_{\mathbf{x},\mathbf{x'}}
\prod\limits_{a=1}^{q}\delta_{i_a,i_a'}\delta_{j_a,j_a'}$,
where $i_1(i'_1)<\cdots<i_q(i'_q)$ and $j_1(j'_1)<\cdots<j_q(j'_q)$. Hopping from one lattice point to another, that always flips the color index, is facilitated by  $t_{\alpha\beta}(\mathbf{x}-\mathbf{x'})$ that conserves the SYK flavor of the fermion. By tuning the magnitude and range of these hoppings, a low-energy dispersion $\varepsilon_\pm(\mathbf{k})$ of the form,
\bea\mylabel{eqn:low_energy_disp}
\varepsilon_\pm(\mathbf{k})\propto\pm|\mathbf{k}|^p,
\eea
can be generated. We choose $p\geq d$ to be an integer to obtain a particle-hole symmetric DOS, with an integrable powerlaw singularity, of the form $g(\varepsilon)\sim|\varepsilon|^{-\gamma}$, where $\gamma=1-d/p$. Evidently, $\gamma$ can be varied between 0 to 1 by tuning $p$ and $d$. Such dispersions with a low-energy form given by \eqn{eqn:low_energy_disp} can be easily `designed' for a  $d$-dimensional lattice (see the Supplementary Material (SM), \SMcite{sec:sup:latmod} for details \mycite{SM}), e.g., a lattice dispersion in $d=1$ corresponding to \eqn{eqn:low_energy_disp} is $\varepsilon_\pm(k)\propto \pm|\sin(k/2)|^p$, with $k$ in units of inverse lattice spacing.
The low-energy dispersion in \eqn{eqn:low_energy_disp} implies the following approximate form for the single-particle DOS
\bea\mylabel{eqn:dos}
g(\varepsilon)=g_0|\varepsilon|^{-\gamma}\Theta(\Lambda-|\varepsilon|)
\eea
where $\Theta$ denotes Heaviside step function and  $\Lambda>0$, a energy cutoff, that plays the role of the bandwidth. The constant $g_0=(1-\gamma)/(2\Lambda^{1-\gamma})$ normalizes the integrated DOS to unity. The above low-energy form of the DOS is sufficient to study the low-temperature properties of the model of \eqn{eqn:Ham} for $\Lambda,J\gg T$. 

\paratitle{Saddle-point equations}
The model of \eqn{eqn:Ham} is solvable at the level of saddle point, which becomes exact in the limit $N\to \infty$, i.e., when the SYK dots consist of a large number of SYK flavors. To derive the saddle point equations \mycite{Sachdev2015,Gu2016} we disorder average over the random SYK couplings using replicas and obtain an effective action within a replica-diagonal ansatz in terms of large-$N$ collective field     
\bea\mylabel{eqn:largeNfields}
G_{\alpha\vtr{x}}(\tau_1,\tau_2)=\frac{1}{N}\sum_i\langle c_{i\alpha\mathbf{x}}(\tau_1)c^\dagger_{i\alpha\mathbf{x}}(\tau_2)\rangle,
\eea
and its conjugate $\Sigma_{\alpha\vtr{x}}(\tau_1,\tau_2)$, where $\tau_{1,2}$ denote imaginary time (see \SMcite{sec:sup:disavgact}). Further, retaining color symmetry and lattice translational invariance for the saddle-point, we obtain the following action (per site, per SYK flavor)
\bea\mylabel{eqn:action}
\mathcal{S}=&-\int\dtau_{1,2}\int\de
g(\varepsilon)
\Tr\ln\left[(\partial_{\tau_{1}}+\varepsilon)\delta(\tau_{1}-\tau_{2})+\Sigma(\tau_{1},\tau_{2})\right]\notag\\
&-\int\dtau_{1,2}\left[(-1)^q\frac{J^{2}}{2q}G^{q}(\tau_{1},\tau_{2})G^{q}(\tau_{2},\tau_{1})\right.\notag\\
&\hspace{11em}\left.+\Sigma(\tau_{2},\tau_{1})G(\tau_{1},\tau_{2})\right],
\eea
where $\int d\tau_{1,2}=\int d\tau_1d\tau_2$. The above action leads to self-consistent equations for the collective fields $G$ and $\Sigma$,
\begin{subequations}\mylabel{eqn:SPiwn}
\begin{align}
G(\iwn)=&\int_{-\Lambda}^{\Lambda}\de g(\varepsilon)\left[\iwn-\varepsilon -\Sigma(\iwn)\right]^{-1}\mylabel{eqn:SPiwn_a}\\
\Sigma(\tau)=&(-1)^{q+1}J^2 G^{q}(\tau)G^{q-1}(-\tau)\mylabel{eqn:SPiwn_b},
\end{align}
\end{subequations}
where $\omega_n=2\pi n T$ is the fermionic Matsubara frequency, $T$ is the temperature with $\ci=\sqrt{-1}$, and we have assumed time translation invariance, i.e., $G(\tau_1,\tau_2)=G(\tau_1-\tau_2)$. At the saddle point, $G(\tau)$ is the on-site fermion Green's function and $\Sigma(\tau)$ is the self-energy, which is completely local in this model. The Green's function of the fermion with momentum $\mathbf{k}$ is given by
\begin{align}\mylabel{eqn:Gk}
G_\pm(\mathbf{k},\iwn)=&\left(\iwn-\varepsilon_\pm(\mathbf{k})-\Sigma(\iwn)\right)^{-1}.
\end{align}

\paratitle{Zero-temperature solutions} The low-energy solution of the saddle-point equations \eqref{eqn:SPiwn} can be obtained analytically at $T=0$. At low temperatures, in the limit  $\omega,\ \Sigma(\omega)\ll \Lambda$, we expand the integral for $G(\iwn\to \omega+\ci0^+)$ in \eqn{eqn:SPiwn_a} in powers of $(\omega-\Sigma(\omega))/\Lambda$ to get
\begin{align}\mylabel{eqn:Gzexpan}
G(\omega)\approx & g_0\frac{\pi(1-e^{\ci\gamma\pi})}{\sin(\pi\gamma)\left(\omega-\Sigma(\omega)\right)^{\gamma}}
\end{align}
at the leading order. As we show below, the above equations leads to two possible fixed point solutions -- (1) a new interaction-dominated fixed point for $\omega\ll\Sigma(\omega)$ as $\omega\to 0$ and (2) the original lattice-dominated fixed point for $\omega\gg\Sigma(\omega)$, essentially a `Fermi liquid', where interaction becomes irrelevant for $\omega\to 0$.

In the first case, \eqn{eqn:Gzexpan} becomes
\bea\mylabel{eqn:Gzscenario1}
G(\omega)=g_0(1-e^{\ci\gamma\pi})\pi\csc(\pi\gamma)\Sigma(\omega)^{-\gamma}.
\eea
At zero temperature, the self-consistent solution of eqns.\eqref{eqn:SPiwn_b},\eqref{eqn:Gzscenario1} is obtained by  taking a powerlaw ansatz for $G(\omega)$, as in the conventional SYK model \mycite{Sachdev1993,Sachdev2015}. This leads to
\begin{subequations}
 \begin{align}
 G(\omega)=&C e^{-\ci\theta}\omega^{2\Delta-1}\mylabel{eqn:IntG}\\
 \Sigma(\omega)=&
 \frac{J^2\left(C\Gamma(2\Delta)\sin\theta\right)^{2q-1}}{\Gamma(2\Delta_\Sigma)\sin(\pi\Delta_\Sigma)\pi^{2q-1}}
e^{\ci\pi\Delta_\Sigma}
\omega^{2\Delta_\Sigma-1} \mylabel{eqn:IntSigma}
  \end{align}
  \end{subequations}
 where the fermion scaling dimension $\Delta$ and the prefactor $C$ are determined self-consistently to be(see \SMcite{sec:sup:zeroTsol} for details)
 \begin{align}
 \Delta&=\frac{1+\gamma}{2(1-\gamma+2q\gamma)}\mylabel{eqn:FdimCdef} \\
 C=&\left[\frac{g_{0}\pi^{2\gamma(q-1)+1}}{J^{2\gamma}\cos(\gamma\pi/2)}\left(\frac{\Gamma(2\Delta_{\Sigma})\sin\left(\pi\Delta_{\Sigma}\right)}{\left(\Gamma(2\Delta)\sin(\pi\Delta)\right)^{2q-1}}\right)^{\gamma}\right]^{\frac{2\Delta}{1+\gamma}}\nonumber
 \end{align}
 with $\Delta_\Sigma=(2q-1)\Delta$, and $\Gamma(x)$ is the gamma function. The low-energy saddle-point equations, and the constraint $\Im G(\omega)<0$, completely fix the spectral asymmetry parameter $\theta$ to $\pi\Delta$ in our case, allowing only particle-hole symmetric scaling solutions at the interacting fixed point, which should be contrasted with the usual SYK model where $\theta$ can be tuned by filling.
  
The fact that the fermion dimension $\Delta$ is determined both by the lattice DOS via $\gamma$ and by SYK interactions through $q$, indicates that the fixed point is indeed a `truly' higher-dimensional analogue of the 0-dimensional SYK phase, but yet distinct from it. In fact as $\gamma\rightarrow 1$, $\Delta\rightarrow 1/2q$ and the fermion dimension of the 0-dimensional SYK model is recovered. However, unlike the SYK model which has an asymptotically \emph{exact} infrared time reparametrization symmetry under $\tau \rightarrow f(\tau)$, \eqn{eqn:SPiwn}(b) together with \eqn{eqn:Gzscenario1} are invariant only under time translation and scaling transformations, $\tau\rightarrow a\tau+b$; $a,~b$ being constants. One would expect time reparametrization symmetry to be restored as the 0-dimensional SYK-like fixed point at $\gamma\rightarrow 1$ is approached. 

The dependence of $\Delta$ on $\gamma$(\eqn{eqn:FdimCdef}) also implies that in principle $\Delta$ can be changed continuously starting from $1/2q$ to $1/2$, as $\gamma$ is tuned from $1$ to $0$. However, as evident from \eqn{eqn:IntSigma}, the assumption $\omega\ll\Sigma(\omega)$, is only self consistent as long as $2\Delta_\Sigma-1\geq 1$, i.e.
\bea\mylabel{eqn:gammacritdef}
1\ge\gamma\geq \frac{2q-3}{2q-1}\equiv\gamma_c(q).
\eea
 Therefore, below a $q$-dependent critical value $\gamma_c(q)$ the scaling solution \eqref{eqn:IntG} ceases to exist. This brings us to the saddle-point solution for the perturbative fixed point for $\omega\gg\Sigma(\omega)$, henceforth referred to as lattice-Fermi liquid (LFL). In this limit, \eqn{eqn:Gzexpan} reduces to
 \bea\mylabel{eqn:Gzscenario2}
 G(\omega)\approx&g_0(1-e^{\ci\gamma\pi})\pi\csc(\pi\gamma)(\omega)^{-\gamma}.
\eea 
It can be shown that at this fixed-point $\Sigma(\omega)\sim (J^2/\Lambda)(\omega/\Lambda)^{(1-\gamma)(2q-1)-1}\ll \omega$ for $\gamma<\gamma_c$, i.e., interaction is irrelevant and its effect is only perturbative in $J$ for $\omega\to 0$. The dominant term in the Green's function above is determined only by the singularity in the single-particle DOS and is temperature independent. At finite temperatures there are small corrections of  $O(J^2)$. We use this fact in the next section to numerically verify the existence of the LFL. 

In gist, we find that the system undergoes a quantum phase transition at $\gamma=\gamma_c(q)$ upon increasing $\gamma$ from 0 to 1. For $\gamma<\gamma_c$, we have a LFL, while for $\gamma>\gamma_c$ we get a line of interaction dominated NFLs. This is also indicated by the dynamical exponent, deduced from \eqn{eqn:Gk}, that changes from $z=p$(when $\gamma<\gamma_c$) to $z=p/(2\Delta_\Sigma-1)$(when $\gamma>\gamma_c$) across $\gamma_c$.

\ifdefined\showfigures
\begin{figure}
\includegraphics[width=\figwidth]{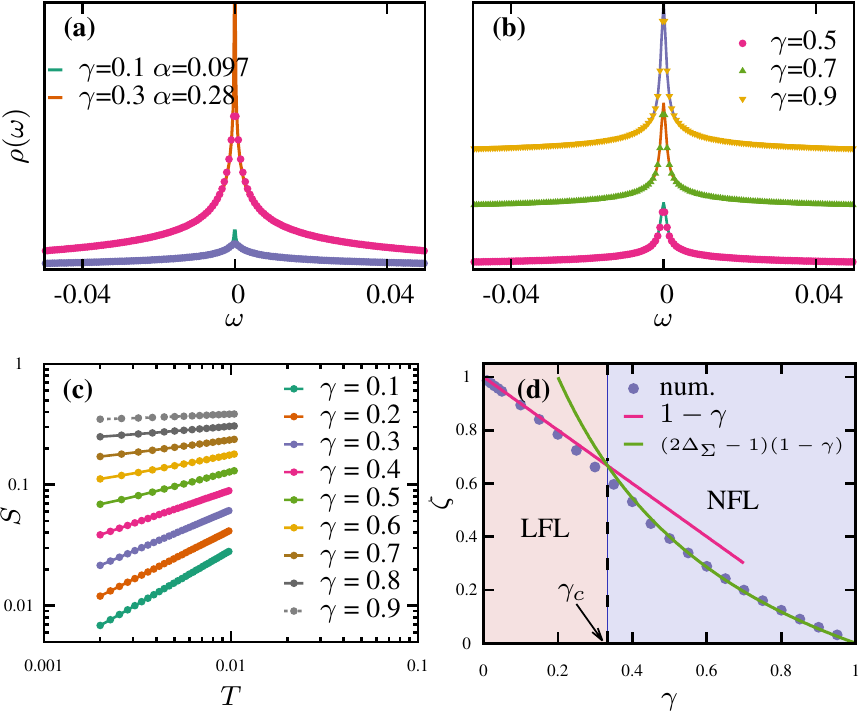}
\caption{{\bf Numerical results ($q=2$, $J=\Lambda=1$):} {\bf (a)} Spectral function at $T=5\times 10^{-3}$ from numerics(points) fitted with the powerlaw $|\omega|^{-\alpha}$(solid lines), where $\alpha=0.097\pm 0.0003,0.28\pm 0.0006$ for $\gamma=0.1,0.3$ respectively. {\bf(b)} Numerical spectral function(points)  at $T=5\times 10^{-3}$ for $\gamma=0.5,0.7,0.9$ compared, \emph{with no fitting parameters}, to those obtained analytically(solid lines) using reparametrization invariance(\eqn{eqn:FiniteTrho}). {\bf(c)}Entropy($S$) vs. temperature($T$) for various $\gamma$ shown in log-log scale, demonstrating the powerlaw dependence of $S$ on $T$. {\bf(d)}The temperature exponent($\zeta$) for $S$(points) as a function of $\gamma$, compared with theoretical predictions(lines) in the lattice-Fermi liquid($\gamma<\gamma_c(=1/3)$)(\eqn{eq:zetaLFL}) and non-Fermi liquid($\gamma>\gamma_c$)(\eqn{eq:zetaNFL}) regions.}
\mylabel{fig:figure2}
\end{figure}
\fi 

\paratitle{Finite-temperature numerics} To gain insight about the finite-temperature properties of the model \eqref{eqn:Ham}, we solve the saddle-point equations \eqref{eqn:SPiwn} numerically at non-zero $T$ for $\Lambda=J=1$ and $q=2$(see \SMcite{sec:sup:numerics} for details). First we calculate the spectral function $\rho(\omega)=-(1/\pi)\Im G(\omega)$ for various $\gamma<\gamma_c(q)$ as shown in \fig{fig:figure2}{\bf(a)}. In this regime $G$ is given by \eqn{eqn:Gzscenario2} and we expect the spectral functions to exhibit the powerlaw singularity in the DOS with small finite-temperature corrections. Therefore we fit the numerically obtained $\rho(\omega)$ with a powerlaw $|\omega|^{-\alpha}$ and find $\alpha\approx\gamma$ (see \fig{fig:figure2}{\bf(a)}), proving the existence of the expected LFL phase. 

The numerical verification of the interacting fixed point for $1>\gamma>\gamma_c(q)$ becomes less straightforward, as unlike the usual SYK model, our system does not possess asymptotically exact reparameterization invariance in the infrared. Still, expecting time reparametrization invariance to be approximately present when $\gamma$ is close to 1, we derive a finite-$T$ expression for the spectral function, by the mapping $\tau=(\beta/\pi)\tan(\sigma\pi/\beta)$ \mycite{Sachdev2015}, to get(see \SMcite{sec:sup:finiteTana}), 
\bea
\rho(\omega)=\frac{C\sin(\pi\Delta)\cosh\left(\beta\omega/2\right)}{\pi^{2}\left(2\pi/\beta\right)^{2\Delta-1}}
{\small\Gamma\left(\Delta-\ci\frac{\beta\omega}{2\pi}\right)\Gamma\left(\Delta+\ci\frac{\beta\omega}{2\pi}\right)},\notag\\
\mylabel{eqn:FiniteTrho}
\eea
where $\beta=T^{-1}$. We compare the above with the $\rho(\omega)$ obtained numerically, anticipating only a qualitative match. Surprisingly, we find an excellent quantitative agreement between the two \emph{without} using any fitting parameter, as demonstrated in  \fig{fig:figure2}{\bf(b)}. Moreover, \eqn{eqn:FiniteTrho} accurately matches the numerical result even for values of $\gamma$ far away from 1 (see \fig{fig:figure2}{\bf(b)}). This not only confirms the existence of NFL fixed points (see \eqn{eqn:IntG}) but also points to an \emph{approximate} emergent reparametrization symmetry at these fixed-points.

In order to verify this result further, we estimate finite-temperature entropy $S(T)$ using $\rho(\omega)$ from \eqn{eqn:FiniteTrho} and a similar finite-temperature form for the self-energy $\Sigma(\omega)$, both of which satisfy a scaling relation, e.g., $\rho(\omega)\sim(T/J)^{2\Delta-1}f(\omega/T)$.  We obtain the entropy via $S=-\partial F/\partial T$, where the free-energy $F$ is obtained by evaluating the action of \eqn{eqn:action} with the scaling form for $\rho(\omega)$ and $\Sigma(\omega)$(see \SMcite{sec:sup:thermo} for details, and also \mycite{Parcollet1998}). We find that the low-temperature entropy vanishes with a power law as $T\to 0$, i.e. $S\sim T^\zeta$, where the exponent
\bea\mylabel{eq:zetaNFL}
\zeta=(2\Delta_\Sigma-1)(1-\gamma)
\eea
 varies between $2/(2q-1)$ and $0$ for $\gamma_c\leq\gamma<1$. We have also verified that as $\gamma\rightarrow 1$, $S$ recovers the usual non-zero value at $T=0$ for the 0-dimensional SYK model (see \SMcite{sec:sup:thermo}, \eqn{eq:sup:SlimitSYK}). 
 
 To test the asymptotic form $S\sim T^\zeta$ , we numerically evaluate entropy from the free-energy for a range of $\gamma$ values and find that the entropy indeed varies as a powerlaw with $T$ (see \fig{fig:figure2}{\bf(c)}). We extract the power of $T$ from the slope of $\ln(S)$ vs. $\ln(T)$ and plot it as a function of $\gamma$ in \fig{fig:figure2}{\bf(d)}, along with the analytical estimate for $\zeta$ (represented by a green line). Again for $\gamma>\gamma_c$, we find a remarkable match of the numerical values with our theoretical predictions, confirming the approximate time reparametrization symmetry of the saddle-point equations \eqref{eqn:SPiwn}, as indicated by the spectral function calculation.

 For the LFL ($0\leq\gamma<\gamma_c$), the expected entropy-temperature relationship is given by $S\sim T^{1-\gamma}$ hence the exponent $\zeta$ becomes
\bea\mylabel{eq:zetaLFL}
\zeta=1-\gamma.
\eea
 The temperature exponent, represented by a red line in \fig{fig:figure2}{\bf(d)}, is a linear function of $\gamma$, and matches exactly with the numerically obtained one around $\gamma=0$.
The most notable point is the abrupt change in the entropy exponent $\zeta$ at $\gamma_c$. For $\gamma>\gamma_c$ the numerically obtained exponent closely follows that of the NFL phase, providing a clear indication of the underlying zero-temperature phase transition at $\gamma_c$.  

\ifdefined\showfigures
\begin{figure}[t!]
\includegraphics[width=\figwidth]{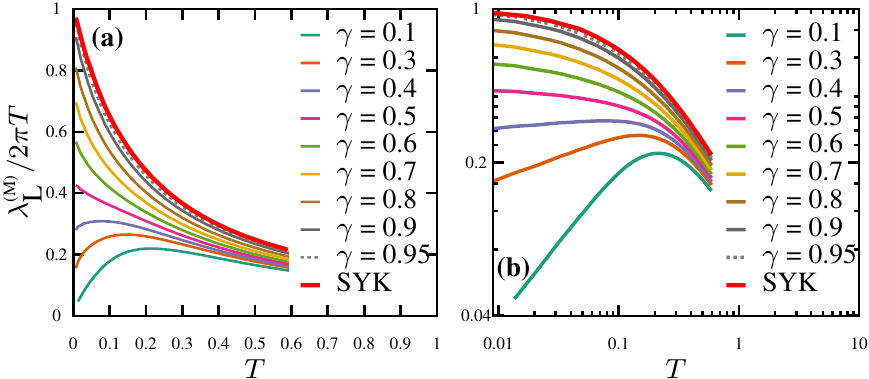}
\caption{{\bf Zero-mode Lyapunov exponent($\lmax$) for $q=2$, $J=\Lambda=1$.} {\bf(a)}  $\lmax/2\pi T$ vs temperature($T$) for $\gamma=0.1-0.95$, bold red line is for 0-dim SYK. {\bf(b)} Same plot in log-log scale, showing upclose the change from a chaotic to a non-chaotic fixed-point.} 
\mylabel{fig:figure3}
\end{figure}
\fi 
 
\paratitle{Chaos and thermalization} While the spectral and thermodynamic quantities for our system clearly indicate that the NFL obtained in the regime $\gamma_c<\gamma<1$ is distinct from the usual SYK model, it is particularly interesting to explore the distinction further in terms of quantum chaos or information scrambling that gives an early-time diagnostic of thermalization \mycite{Gu2017}. The SYK model is known to be the most efficient scrambler like a black hole \mycite{KitaevKITP}, namely the Lyapunov exponent $\lambda_L$, characterizing decay of a typical out-of-time-ordered (OTO) correlator, e.g., $\langle \cd_i(t) \cd_j(0) c_i(t) c_j(0) \rangle
\simeq f_0-(f_1/N)e^{\lambda_L t}+\mathcal{O}\left(N^{-2}\right)$,
saturates the upper bound, $\lambda_L=2\pi T$, imposed by quantum mechanics \mycite{Maldacena2015,BanerjeeAltman2016}. A natural question is then to ask, whether our new NFL states behave similarly or differently. To this end, we generalize the OTO correlator for our two-band lattice system as
\bea\mylabel{eqn:OTOCs}
\frac{1}{N^2}\sum_{ij}\langle \cd_{i\alpha x}(t) \cd_{j\beta x'}(0) c_{i\alpha x}(t) c_{j\beta x'}(0) \rangle
\notag\\
\frac{1}{N^2}\sum_{ij}\langle c_{i\alpha x}(t) \cd_{j\beta x'}(0) \cd_{i\alpha x}(t) c_{j\beta x'}(0) \rangle \notag
\eea
and find that their time evolution is governed by two lattice-momentum ($\mathbf{q}$) dependent modes, an intraband mode and an interband one. At $\mathbf{q}=\mathbf{0}$ the lattice dispersion enters into the expressions of OTO correlator for both the modes  only through the overall DOS $g(\varepsilon)$  (see \SMcite{sec:sup:lyapexp}, \eqn{eq:sup:Kzero_final}).  Also at $\mathbf{q}=\mathbf{0}$ the intraband  mode has the larger Lyapunov exponent ($\lmax$) among the two and therefore dominates the onset of chaos (see \SMcite{sec:sup:lyapexp}). 

We numerically  calculate $\lmax$ for $q=2$ as a function of $T$ for various values of $\gamma$ on both sides of the quantum critical point $\gamma_c(q=2)=1/3$ as show in \fig{fig:figure3}. We find that $\lmax/2\pi T$ becomes identical with that of 0-dimensional SYK model (represented by a bold red line in \fig{fig:figure3}{\bf(a)}) as $\gamma\rightarrow 1$. In particular $\lmax/2\pi T\rightarrow 1$ as $T\rightarrow 0$, thereby saturating the chaos bound. 

For $\gamma_c<\gamma<1$, our numerical results indicate in the limit $T\rightarrow 0$, $\lmax/2\pi T\rightarrow \alpha$, where $0<\alpha<1$. This distinct behavior from original SYK model implies that the NFL fixed points here \emph{do not} saturate the chaos bound although they are still very efficient scramblers with Lyapunov exponent $\propto T$ at low temperature. Interestingly  similar behavior has been reported \mycite{Patel2016}, in systems, albeit in a less controlled calculation, involving fermions coupled to a gauge field. Around the neighborhood of $\gamma_c$ for $\gamma=(0.3-0.5)$, $\alpha$ starts to turn around and tend towards zero (see \fig{fig:figure3}{\bf(b)}), signifying a change in the chaotic behavior of the system. Finally, for $\gamma<\gamma_c$, $\lmax/2\pi T\rightarrow 0$ as $T\rightarrow 0$ indicating the presence of a slow scrambling phase similar to a Fermi-liquid.

\paratitle{Discussion}
In this paper, we have developed a model that achieves a higher-dimensional generalization of the SYK model. The class of NFL phases discovered here should provide a platform to study transport properties in strongly interacting quasiparticle-less lattice systems, with non-random hoping amplitudes. The latter allows to go beyond purely diffusive transport \mycite{Gu2016,SJian2017b,CJian2017,Song2017} and study the interplay of fermion dispersion and interaction in NFL phases, in a manner not possible prior to this work. To this end, it would be interesting to explore connection between transport and scrambling in our model \mycite{Arijit2017} and contrast with that in other lattice generalizations of SYK model with random hoppings \mycite{Gu2016}. This work also has relevance towards the study of interaction effects near Lifshitz transitions -- a question that is relevant from the perspective of transitions between band insulating topological phases that are generically separated by such transitions. Finally, it would be interesting to understand  how does approximate time reparametrization symmetry emerges in our system and what implications does it have from the point of view of the gravitational dual.

\paratitle{Acknowledgements} The authors AH and VBS thank DST, India for support.

\bibliography{SYK}

\def\makeSM{1}
\ifdefined\makeSM
\clearpage
\newpage

\appendix
\renewcommand{\appendixname}{}
\renewcommand{\thesection}{{S\arabic{section}}}
\renewcommand{\theequation}{\thesection.\arabic{equation}}
\setcounter{page}{1}

\widetext

\centerline{\bf Supplemental Material}
\centerline{\bf for}
\centerline{\bf \titlename}
\centerline{by Arijit Haldar, Sumilan Banerjee and Vijay B.~Shenoy}
\author{Arijit Haldar}
\email{arijit@physics.iisc.ernet.in}
\author{Vijay B. Shenoy}
\email{shenoy@physics.iisc.ernet.in}
\affiliation{Centre for Condensed Matter Theory, Department of Physics, Indian Institute of Science, Bangalore 560 012, India}
\section{Lattice-models and dispersions}\mylabel{sec:sup:latmod}
In this section we outline a procedure to obtain lattice dispersions with the low-energy form given by
\begin{equation}\mylabel{sup:eq:low_energy_disp}
\varepsilon(\vk)=\pm|\vk|^{p},
\end{equation}
where $\vk=\{k_{1},\cdots,k_{d}\}$ is a $d-$dimensional vector in the Brillouin zone. These kind of particle-hole symmetric dispersions can be obtained from Hamiltonians with the following form
\begin{align}
\sum_{k}H(k)=\sum_{k}
\left[
\begin{array}{cc}
\cd_{R}(\vk) & \cd_{B}(\vk)
\end{array}
\right]
\left[
\begin{array}{cc}
0 & \hat{t}(\vk)\\
\hat{t}(\vk)^{*} & 0
\end{array}
\right]
\left[
\begin{array}{c}
c_{R}(\vk)\\
c_{B}(\vk)
\end{array}
\right],
\label{eq:block_diagonal}
\end{align}
where $R/B$ denotes the fermion color indices red/blue and $\hat{t}(\vk)$ is a function of lattice-momentum $\vk$ obtained by a Fourier transform of the hopping-amplitudes $t_\vtr{r}$ which are yet to be determined, such that
\bea
\hat{t}(\vk)=\sum_{\vtr{r}}-t_{\vtr{r}}e^{-\ci \vtr{r}\cdot\vk},
\eea
 The hopping $t_\vtr{r}$ is the amplitude for a red color fermion to hop to a site separated by $\vtr{r}$ and flip its color to blue. The resulting two-band ($\pm$) energy dispersion is given by
\begin{equation}
\varepsilon_{\pm}(\vk)=\pm\sqrt{\hat{t}(\vk)\hat{t}(\vk)^{*}}.
\end{equation}
A natural candidate for the dispersion whose low energy form is given by \eqn{sup:eq:low_energy_disp} would be
\bea\mylabel{eq:sup:high_energy_disp}
\varepsilon_\pm(\vk)=
\left(\sum_{i=1}^{d}\sin(k_{i}/2)^{2}\right)^{p/2}
\eea
since near $\vk=\vtr{0}$ we have
$
\left(\sum\limits_{i=1}^{d}\sin(k_{i}/2)^{2}\right)^{p/2}
\approx
\left(\sum\limits_{i=1}^{d}(k_{i}/2)^{2}\right)^{p/2}\propto\pm|\vk|^p
$. We choose $k_i/2$ instead of $k_i$ since we want the dispersions to be gapless only at $\vk=\vtr{0}$. The problem of finding a suitable $\hat{t}(\vk)$ then reduces to factorizing the following equation
\bea\mylabel{eq:sup:tktk}
\hat{t}(\vk)\hat{t}(\vk)^*=\left(\sum_{i=1}^{d}\sin(k_{i}/2)^{2}\right)^p.
\eea
When $p$ is an even integer we can factorize the RHS  of the above equation by taking
\begin{equation}
\hat{t}(\vk)=\left(\sum_{i=1}^{d}\sin(k_{i}/2)^{2}\right)^{p/2}.
\end{equation}
Then by substituting $\sin(k_{i}/2)=(e^{\ci k_{i}/2}-e^{-\ci k_{i}/2})/2\ci$ and expanding the resulting expression we can easily read off the hopping amplitudes from the coefficients in the expansion. For e.g. for dimension $d=1$ and $p=2$ we get
\bea
\hat{t}(\vk)=\sin(k/2)^2=\frac{1}{2}-\frac{1}{4}e^{\ci k}-\frac{1}{4}e^{-\ci k},
\eea
implying a hopping configuration which is shown in \fig{fig:sup:figure4}{\bf(a)}. Lattices constructed in this way will generically have inversion ($x\rightarrow-x$) and color interchange ($R\leftrightarrow B$) symmetries. When $p$ is an odd integer the function $\hat{t}(\vk)$ that needs to be chosen for factorizing the RHS of \eqn{eq:sup:tktk} is a bit more subtle. For e.g. when dimension $d=1$ 
\bea
\hat{t}(\vk)=-\ci e^{\ci\frac{k}{2}}\sin(k/2)^{p}
\eea
turns out to be the correct choice. This can then be expanded as done before for the $p$ even case to obtain the hopping amplitudes, e.g., for $p=3$, $\hat{t}(\vk)$ is given by
\bea
\hat{t}(\vk)=-\ci e^{\ci\frac{k}{2}}\sin(k/2){}^{p=3}=\frac{3}{8}-\frac{3}{8}e^{\ci k}-\frac{1}{8}e^{-\ci k}+\frac{1}{8}e^{\ci2k},
 \eea 
 \ifdefined\showfigures
\begin{figure}
\includegraphics[scale=1.0]{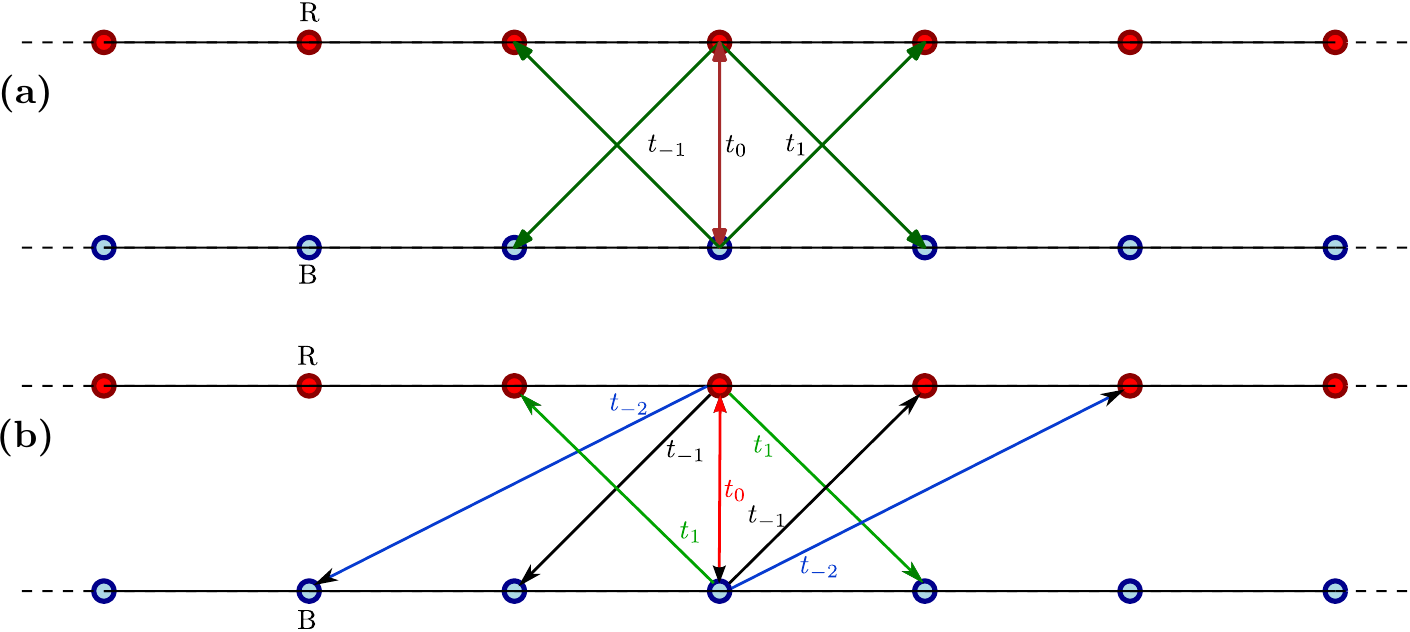}
\caption{{\bf Hopping configurations:} {\bf(a)} $p=2$ (even) case -- The hopping amplitudes are $t_0=-1/2$, $t_1=t_{-1}=1/4$ and generically we have $t_{-\vtr{r}}=t_{\vtr{r}}$ with $p/2=1$  nearest neighbors needed to left and right of the source site. The lattice has inversion and color interchange symmetries separately. {\bf(b)} $p=3$ (odd) case -- The amplitudes are given by $t_0=-3/8$, $t_1=1/8$, $t_{-1}=3/8$ and $t_{-2}=-1/8$. Generically $t_{-r}\neq t_{r}$ , also $(p-1)/2 =1$ neighbors to the right and $(p+1)/2=2$ neighbors to the left are needed for $R\rightarrow B$ hops. Note that the lattice does not have inversion and color interchange symmetry separately, however the product of the two symmetry operations is still a symmetry.}\mylabel{fig:sup:figure4}
\end{figure}
\fi 
implying a hopping structure as shown in \fig{fig:sup:figure4}{\bf(b)}. The resultant lattice in this situation is qualitatively different from the $p$ even case(see \fig{fig:sup:figure4}{\bf(a)}), in that it is no longer invariant under inversion and color interchange operations. However the product of the two operations still remains a symmetry for this kind of a lattice. \section{Disorder averaged action and saddle-point equations}\mylabel{sec:sup:disavgact}
We use the Hamiltonian for our model given by eqn.~(1) of the main text
\bea\mylabel{eqn:sup:Ham}
\cH =-\sum_{i\alpha\alpha'\vtr{x}\vtr{x'}}t_{\alpha\alpha'}\left(\vtr{x}-\vtr{x'}\right)c^\dagger_{i\alpha\vtr{x}}c_{i\alpha'\vtr{x'}}
-\mu\sum_{i\alpha\vtr{x}}
c^\dagger_{i\alpha\vtr{x}}c_{i\alpha\vtr{x}}
+\sum_{\alpha\vtr{x}}\ \ \sum\limits_{\mathclap{\substack{i_1,\cdots,i_{q},\\j_1,\cdots,j_{q}}}}  J^{\alpha,\vtr{x}}_{i_1,\cdots,i_{q};j_1,\cdots,j_{q}}c^\dagger_{i_q\alpha\vtr{x}}\cdots c^\dagger_{i_1\alpha\vtr{x}} c_{j_1\alpha\vtr{x}} \cdots c_{j_q\alpha\vtr{x}},\notag\\
\eea
to write down the Euclidean-time action in terms of Grassman variables $(\cb,c)$
\begin{align*}
\act= & \int_0^\beta d\tau\left[\sum_{i\alpha \vtr{x}}\bar{c}_{i\alpha\vtr{x}}\partial_{\tau}c_{i\alpha\vtr{x}}
+\cH(\cb,c)\right],
\end{align*}
where $\beta=T^{-1}$. The $J$s are independent complex-random variables, chosen from a Gaussian distribution with variance $\overline{|J_{i_{1}\dots i_{q}j_{1}\dots j_{q},x}|^{2}}=J^{2}/qN^{2q-1}(q!)^2$, and properly antisymmetrized. The disorder averaged action can then be written down using the replica-trick as 
 \begin{align}
 \act=&\int\dtau_{1}\int\dtau_{2}\left[\sum_{i\vtr{x}\vtr{x'}\alpha\alpha',r_1}\cb_{i\alpha xr_1}(\tau_1)
 \left[(\partial_{\tau_{1}}-\mu)\delta_{\alpha\alpha'}\delta_{\vtr{x}\vtr{x'}}
 +t_{\alpha\alpha'}\left(\vtr{x}-\vtr{x'}\right)\right]
 \delta(\tau_{1}-\tau_{2})c_{i\alpha' \vtr{x'}r_1}(\tau_{2})\right.\notag\\
&\left.-(-1)^q\frac{J^2}{2qN^{2q-1}} \sum_{\alpha\vtr{x}r_1r_2}\left(\sum_{ij}\cb_{i\alpha \vtr{x}r_1}(\tau_1)c_{i\alpha\vtr{x}r_2}(\tau_2)\cb_{i\alpha \vtr{x}r_2}(\tau_2)c_{i\alpha\vtr{x}r_1}(\tau_1)\right)^q\right]
\end{align}
where $r_1,r_2$ are the replica indices. We introduce the large-$N$ field $G_{\alpha\vtr{x}r_1r_2}(\tau_{1},\tau_{2})=\frac{1}{N}\sum\limits_{i}\cb_{i\alpha\vtr{x}r_2}(\tau_{2})c_{i\alpha\vtr{x}r_1}(\tau_{1})$ and the Lagrange multiplier $\Sigma_{\alpha\vtr{x}r_1r_2}(\tau_1,\tau_2)$ to obtain
\begin{align}
 S=&\int\dtau_{1}\int\dtau_{2}\left[\sum_{i\vtr{x}\vtr{x'}\alpha\alpha',r_1}\cb_{i\alpha\vtr{x}r_1}(\tau_1)
 \left[(\partial_{\tau_{1}}-\mu)\delta_{\alpha\alpha'}\delta_{\vtr{x}\vtr{x'}}
 +t_{\alpha\alpha'}\left(\vtr{x}-\vtr{x'}\right)\right]
 \delta(\tau_{1}-\tau_{2})
 c_{i\alpha' \vtr{x'}r_1}(\tau_{2})\right.\notag\\
& \left.-N\sum_{\alpha\vtr{x}r_1r_2}\left[(-1)^q\frac{J^{2}}{2q}
G_{\alpha\vtr{x}r_1r_2}^{q}(\tau_{1},\tau_{2})G_{\alpha\vtr{x}r_2r_1}^{q}(\tau_{2},\tau_{1})+G_{\alpha\vtr{x}r_1r_2}(\tau_{1},\tau_{2})\Sigma_{\alpha\vtr{x}r_2r_1}(\tau_{2},\tau_{1})\right]\right].
 \end{align}
The kinetic part of the action can be diagonalized by transforming the real space fermions $\cb_{i\alpha\vtr{x}r_1}(\tau)$ to Bloch fermions, $\db_{ib\vk r_1}(\tau)$
, where $b$ denotes the band index and takes the value $+$($-$) for the upper(lower) band. Next we impose lattice translational and fermion color interchange invariance, such that $\Sigma_{\alpha\vtr{x}}=\Sigma$ and $G_{\alpha\vtr{x}}=G$, then set $\mu=0$ and use the replica-diagonal ansatz to obtain
\begin{align}
\mathcal{S}\equiv\frac{\act}{NN_L} & = \frac{1}{N_L} \int\dtau_{1}\int\dtau_{2}\left[\sum_{b=\pm,\vk}\db_{ib\vk}(\tau_{1})\left[(\partial_{\tau_{1}}+\ek)
\delta(\tau_{1}-\tau_{2})+\Sigma(\tau_{1},\tau_{2})\right]d_{ib\vk}(\tau_{2})\right.\nonumber \\
 & \left.- (-1)^q \frac{J^{2}}{2q}G^{q}(\tau_{1},\tau_{2})G^{q}(\tau_{2},\tau_{1})-\Sigma(\tau_{2},\tau_{1})G(\tau_{1},\tau_{2})\right],
\end{align}
where $N_L$ is twice the total number of lattice sites, including the two colors  and $\ek$ is the band dispersion.
Tracing over the fermionic degrees of freedom we get
\begin{eqnarray}
\mathcal{S}&= & \int\dtau_{1}\int\dtau_{2}\left[-\sum_{\alpha,k}\Tr\ln\left[(\partial_{\tau_{1}}+\ek)\delta(\tau_{1}-\tau_{2})+\Sigma(\tau_{1},\tau_{2})\right] - (-1)^q \frac{J^{2}}{2q}G^{q}(\tau_{1},\tau_{2})G^{q}(\tau_{2},\tau_{1})-\Sigma(\tau_{2},\tau_{1})G(\tau_{1},\tau_{2})\right]\notag\\
\end{eqnarray}
By extremizing the action $\mathcal{S}$,  we obtain the saddle-point equations for $G$ and $\Sigma$ as
\begin{subequations}
\begin{align}\mylabel{eq:sup:SPeqtau12}
\Sigma(\tau_1,\tau_2)= & (-1)^{q+1}J^{2}G^{q-1}(\tau_2,\tau_1)G^{q}(\tau_1,\tau_2)\\
G(\tau_1,\tau_2)= & \langle\bar{c}_{x}(\tau_2)c_{x}(\tau_1)\rangle=\frac{1}{N_L}\sum_{b,\mathbf{k}}G_{b\mathbf{k}}(\tau_1,\tau_2)
\end{align}
\end{subequations}
where 
\bea
G_{b\mathbf{k}}^{-1}(\tau_1,\tau_2)
=-\left[(\partial_{\tau_1}+\ek)\delta(\tau_1-\tau_2)+\Sigma(\tau_1,\tau_2)\right].
\eea
Finally, from the above,  we obtain the saddle-point equations (6) and the corresponding action of eqn.~(5) in the main text for a time-translationally invariant solution.


\section{Zero temperature solutions}\mylabel{sec:sup:zeroTsol}
The saddle-point equations [eqns. (6), main text] can be solved analytically for $T=0$, by taking a powerlaw ansatz for $G$ and $\Sigma$. Analytically continuing $G(\ci\omega_n)$ to complex plane and using $g(\varepsilon)=g_0|\varepsilon|^{-\gamma}$, we obtain
\begin{align}\mylabel{eq:sup:Gzbeta}
G(z)= & g_{0}\int_{-\Lambda}^{\Lambda}\de\frac{|\varepsilon|^{-\gamma}}{\tilde{z}-\varepsilon}=g_{0}\tilde{z}^{-\gamma}\left[\int_{0}^{\frac{\Lambda}{\tilde{z}}}d\varepsilon\frac{\varepsilon{}^{-\gamma}}{1-\varepsilon}+\int_{0}^{\frac{\Lambda}{\tilde{z}}}d\varepsilon\frac{\varepsilon{}^{-\gamma}}{1+\varepsilon}\right]
\end{align}
where $\tilde{z}=z-\Sigma(z)$ and $g_0=(1-\gamma)/2\Lambda^{1-\gamma}$. The above integrals can be written in terms of the incomplete beta function, $B(z;a,b)=\int_0^z u^{a-1}(1-u)^{b-1}\textup{d}u$,
and are obtained for $|\tilde{z}|\to0$ as
\begin{align*}
\int_{0}^{\frac{\Lambda}{\tilde{z}}}\frac{\varepsilon{}^{-\gamma}}{1-\varepsilon}= & B\left(\frac{\Lambda}{\tilde{z}};1-\gamma,0\right)\approx(-1)^{1+\gamma}\pi\csc(\gamma\pi)+\frac{1}{\gamma}\left(\frac{\tilde{z}}{\Lambda}\right)^{\gamma}+\frac{1}{1+\gamma}\left(\frac{\tilde{z}}{\Lambda}\right)^{\gamma+1}+\dots\\
\int_{0}^{\frac{\Lambda}{\tilde{z}}}\frac{\varepsilon{}^{-\gamma}}{1+\varepsilon}= & (-1)^{1+\gamma}B\left(-\frac{\Lambda}{\tilde{z}};1-\gamma,0\right)\approx\pi\csc(\gamma\pi)-\frac{1}{\gamma}\left(\frac{\tilde{z}}{\Lambda}\right)^{\gamma}+\frac{1}{1+\gamma}\left(\frac{\tilde{z}}{\Lambda}\right)^{\gamma+1}+\dots
\end{align*}
so that
\begin{align}
G(z)\approx & g_{0}\tilde{z}^{-\gamma}\left[(1+(-1)^{1+\gamma})\pi\csc(\gamma\pi)+\frac{2}{1+\gamma}\left(\frac{\tilde{z}}{\Lambda}\right)^{\gamma+1}+\dots\right]\label{eq:sup:Gz}.
\end{align}
The leading order term of the above equation leads to the low-energy saddle-point equation (8) of the main text via analytical continuation $z\to \omega+\ci \eta$.

\paragraph{\underline{Case -- Non-Fermi liquid(\NFL) limit:}} We obtain the self-consistency condition [eqn.~(9), main text] for the NFL fixed points by droping  $\omega$ in $\tilde{z}=\omega-\Sigma(\omega)$.
By taking the powerlaw ansatz of eqn.~(10a) (main text), we obtain $T=0$ spectral function,
\begin{align}\mylabel{eq:sup:specfNFL}
\rho(\omega)= & -\frac{1}{\pi}\Im G(\omega)
=\frac{1}{\pi}
\begin{cases}
\sin\phi\frac{C}{(\omega)^{\alpha}}&\omega>0\\
\sin(\phi+\alpha\pi)\frac{C}{(-\omega)^{\alpha}}&\omega<0.
\end{cases}
\end{align}
Since $\rho(\omega)>0$ we must have $0<\phi<\pi(1-\alpha)$ and particle-hole
symmetry, $\rho(\omega)=\rho(-\omega)$, implies $\sin\phi=\sin(\phi+\alpha\pi)$. By using the spectral representation, 
\begin{align}
G(\tau)= &
-\int_{-\infty}^{\infty}d\omega
\begin{cases}
\frac{\rho(\omega)}{e^{-\beta\omega}+1}e^{-\omega\tau}&\tau>0\\
\frac{\rho(\omega)}{e^{\beta\omega}+1}e^{-\omega\tau}&\tau<0\\
\end{cases}= \frac{C}{\pi}\sin\phi\Gamma(1-\alpha)\frac{-\mathrm{sgn}(\tau)}{|\tau|^{2\Delta}},\\
\end{align}
where the fermion scaling dimension $\Delta$ is obtained from $2\Delta=1-\alpha$. Substituting the above into eqn.~(6b) of the main text, the self-energy is obtained as
\begin{align}\mylabel{eq:sup:SEtauNFL}
\Sigma(\tau)=J^{2} \frac{C^{2q-1}}{\pi^{2q-1}}\Gamma^{2q-1}(2\Delta)\sin^{2q-1}(\phi)\frac{-\mathrm{sgn}(\tau)}{|\tau|^{2\Delta_\Sigma}}, &\ \ \  \Delta_\Sigma=(2q-1)2\Delta,
\end{align}
which leads to
\begin{align}\mylabel{eq:sup:SERwNFL}
\Sigma(\omega)=J^{2}\pi  \frac{C^{2q-1}}{\pi^{2q-2}}\Gamma^{2q-1}(2\Delta)\frac{\sin^{2q-1}(\phi)}{\Gamma(2\Delta_\Sigma)\sin(\pi\Delta_\Sigma)} e^{-i\pi\Delta_\Sigma}\omega^{2\Delta_\Sigma-1},
\end{align}
The self-consistency condition [eqn.~(9), main text] fixes $C$ to be
\begin{align}
C= & \left[\frac{g_{0}\pi}{\cos(\gamma\pi/2)}J^{-2\gamma}\left(\frac{\pi}{\Gamma(2\Delta)\sin(\pi\Delta)}\right)^{\gamma(2q-1)}\left(\frac{\Gamma(2\Delta_\Sigma)\sin(\pi\Delta_\Sigma)}{\pi}\right)^{\gamma}\right]^{\frac{2\Delta}{1+\gamma}}
\end{align}
and the fermion dimension $\Delta$ to
\begin{align}\mylabel{eq:sup:DeltaNFL}
\Delta= & \frac{1+\gamma}{2(1-\gamma+2\gamma q)},
\end{align}
which is valid as long as $\omega\ll\Sigma(\omega)\propto\omega^{2(2q-1)\Delta-1}$
as $\omega\to0$. This implies  $2(2q-1)\Delta-1\leq 1$ in order for \eqn{eq:sup:DeltaNFL} to hold, leading to a critical value of $\gamma$, namely 
\begin{align}\mylabel{eq:sup:gammac}
\gamma_{c}(q)= & \frac{2q-3}{2q-1}.
\end{align}
$\gamma_c\to 1$ as $q\to\infty$ and the regime for NFL shrinks to a point at $\gamma=1$. Therefore fermion scaling dimension lies in the range,
\begin{align}\mylabel{eq:sup:scalingdimlimNFL}
\frac{1}{2q} & \leq\Delta\leq\frac{1}{2q-1}.
\end{align}
Finally, we determine the constraint on the phase $\phi$ to be
\begin{align*}
\phi= \pi\Delta-\frac{\pi\gamma}{2}(1-\mathrm{sgn}(\sin(\pi(2q-1)\Delta)),
\end{align*}
where, for the allowed range of $\Delta$ in \eqn{eq:sup:scalingdimlimNFL}, the second term above is zero implying $\phi= \pi\Delta$, which corresponds to particle-hole symmetry. Hence, the low-energy scaling solution in NFL fixed point pins $\phi$ to the particle-hole symmetric point.

\paragraph{\underline{Case -- Lattice-Fermi liquid(\LFL) limit:}} A different $T=0$ self-consistent solution is obtained from eqn.~(8) (main text) for $\gamma<\gamma_c$. In this regime the effect of interaction contributes perturbatively and the solution can be obtained by neglecting $\Sigma(\omega)$ in $\tilde{z}=\omega-\Sigma(\omega)$ at the leading order in eqn.~(8) to get eqn.~(13) in the main text.
In this case, $\rho(\omega)\propto|\omega|^{-\gamma}$, and the leading order self-energy is evaluated to be
 \bea
 \Sigma(\omega)=\frac{J^2}{\Lambda}\left(\frac{\omega}{\Lambda}\right)^{(1-\gamma)(2q-1)-1}.
 \eea
 Evidently the above self-enrgy correction is irrelevant at this fixed point for $\omega\to0$, since  $\omega^{(1-\gamma)(2q-1)-1}<<\omega$ when $\gamma<\gamma_c(q)$.\section{Numerical solutions}\mylabel{sec:sup:numerics}
We solve the imaginary-time and the real-frequency versions of the saddle-point equations (6) (main text) self-consistently using an iterative scheme, in order to calculate the thermal and spectral quantities. In this section we discuss the numerical details and algorithms that were used to perform the numerical calculations.
\subsection{Solving the imaginary-time version and calculating thermal quantities}
\mylabel{subsec:sup:numtau}
To numerically solve the saddle-point equations [eqns.~(6), main text] we start with an initial guess for the Green's functions $G(\ci\omega_n)$ (which we took to be the non-interacting version $G^0(\ci\omega_n)=(\ci\omega_n+\mu)^{-1}$). Using Fast Fourier Transform(FFT), $G(\tau)$ is obtained by evaluating the Matsubara sum
\bea\mylabel{eqn:sup:Gwn_to_Gtau}
G(\tau)=\frac{1}{\beta}\sum_{\ci\omega_n}G(\ci\omega_n)e^{-\ci\omega_n\tau},
\eea
which is then used to compute the self-energy $\Sigma(\tau)$ from eqn.~(6b) (main text). Finally, the iterative loop is closed by transforming  $\Sigma(\tau)$ to $\Sigma(\ci\omega_n)$ via another FFT to evaluate
\bea
\Sigma(\ci\omega_n)=\int\limits_0^\beta \D{\tau} \Sigma(\tau)e^{\ci\omega_n\tau}
\eea
 and then calculating the new $G(\ci\omega_n)$. This process is carried out till the difference between the new $G(\ci\omega_n)$ and the old one has become desirably small.
 
We mention now a few subtleties that are needed to be taken care off inorder to attain convergence at lower temperatures. First, the use of eqn.~(6a) (main text) in its original form was not sufficient to attain convergence, instead the following was used
\bea
G_{\textup{new}}(\ci\omega_n)=\alpha\ G_{\textup{old}}(\ci\omega_n)+(1-\alpha)
\int_{-\Lambda}^\Lambda
\de\ 
g(\varepsilon)
\left[\ci\omega_n+\mu-\varepsilon-\Sigma_s(\ci\omega_n)\right]^{-1},
\eea 
where we fed back a part of the old $G$ along with the usual expression. In all our numerical calculations we took $\alpha=0.8$. Second, in order to obtain $G(\tau)$ from $G(\ci\omega_n)$ a direct evaluation of Fourier sum in \eqn{eqn:sup:Gwn_to_Gtau} produces too much Gibb's oscillations near the end points of $G(\tau)$. This can be remedied by subtracting out the non-interacting part $1/\ci\omega_n$  from  $G(\ci\omega_n)$  and then taking the Fourier transform, after which we add back the analytical expression for the Fourier transform of the non-interacting part, i.e.,
\bea
G(\tau)=\frac{1}{\beta}\sum\limits_{\ci\omega_n}\left[G(\iwn)-\frac{1}{i\omega_n}\right]e^{-\ci\omega_n\tau}-\frac{1}{2}.
\eea
The reason this works is due to the fact that at large $\omega_n$ ($\iwn>>\varepsilon,\Sigma$), $G(\ci\omega_n)\sim 1/\iwn$, and subtracting out $1/\ci\omega_n$ makes it fall off faster, thereby rendering the FFT more controlled. As a result, discretizing the $\tau$ domain into $2^{17}$ intervals for performing the FFTs were sufficient to attain convergence to temperatures as low as $0.007$. Increasing the discretization further should in principle allow access to even lower temperatures.

 We use the resulting self-consistent $G(\tau)$, $\Sigma(\tau)$ to evaluate the free- energy in the following regularized form
\begin{align}
F =&  -\frac{1}{\beta}\sum\limits_{\ci\omega_n}\int\de g(\varepsilon)\ln\left[\frac{\ci\omega_n+\mu-\varepsilon-\Sigma(\ci\omega_n)}{\ci\omega_n +\mu}\right]  
- \frac{J^2}{2 q}  \int_0^\beta  \D{\tau}   G^q(\beta - \tau) G^q(\tau)  
 + \int_0^\beta \D{\tau} \,  \Sigma(\tau) G(\beta-\tau)\notag\\ 
&  -\frac{1}{\beta}\ln(1+e^{\beta \mu})
\end{align}
Subsequently, the entropy $S=-\partial F/\partial T$ is evaluated by computing numerical derivative of $F(T)$. The regularized form of free energy is required because Matsubara sums of the form $\sum\limits_{\ci\omega_n}\ln[-(\ci\omega_n+\mu-\Sigma(\ci\omega_n))]e^{\ci\omega_n0^+}$
are not numerically convergent. In order to make them convergent we subtract from it the free gas contribution 
\bea\mylabel{eqn:sup:freegas}
\sum\limits_{\ci\omega_n}\ln[-(\ci\omega_n+\mu)]e^{\ci\omega_n0^+}
\eea
and then carry out the sum numerically. Later we add back the analytical expression, $-\ln(1+e^{\beta \mu})$, for \eqn{eqn:sup:freegas}.
\subsection{Solving the real frequency version and calculating spectral functions}
To obtain the spectral functions we follow a similar approach to \cite{BanerjeeAltman2016}, and analytically continue the saddle point equations (6) (main text) to real frequencies, $\ci\omega_n\rightarrow \omega+\ci\eta$. The expression for self-energy is given by
\bea\mylabel{eqn:sup:Sigma_wplus}
\Sigma(\omega^+)
=-\ci\int\limits_0^{\infty}\D{t}\ 
e^{\ci\omega t}
J^2
\left\{
 n_{1}^{q-1}(t)
 n_{2}^{q}(t)
+
n_{3}^{q-1}(t)
n_{4}^{q}(t)
\right\}
\eea
where $n_{(1-4)}(t)$ are defined as
\bea\mylabel{eqn:sup:nint_def}
\begin{array}{ll}
n_{1}(t)=\int\limits_{-\infty}^{+\infty}\D{\Omega}\ \rho(\Omega)n_F(-\Omega) e^{+\ci\Omega t},&
n_{2}(t)=\int\limits_{-\infty}^{+\infty}\D{\Omega}\ \rho(\Omega)n_F(\Omega) e^{-\ci\Omega t}\\
n_{3}(t)=\int\limits_{-\infty}^{+\infty}\D{\Omega}\ \rho(\Omega)n_F(\Omega) e^{+\ci\Omega t},&
n_{4}(t)=\int\limits_{-\infty}^{+\infty}\D{\Omega}\ \rho(\Omega)n_F(-\Omega) e^{-\ci\Omega t}.
\end{array}
\eea
The equation for $G$ [eqn.~(6a), main text] is also analytically continued to real frequency to obtain the retarded function 
\bea\mylabel{eqn:sup:G_w}
G(\omega)=\int\de g(\varepsilon)[\omega+\mu-\varepsilon-\Sigma(\omega)]^{-1}
\eea
 The spectral function is related to the retarded Green's function as
\bea\mylabel{eqn:sup:spectral_func_def}
\rho(\omega)=-\frac{1}{\pi}\Im G(\omega).
\eea
Using \eqn{eqn:sup:Sigma_wplus}, \eqn{eqn:sup:G_w} and \eqn{eqn:sup:spectral_func_def} we iteratively solve for the spectral function $\rho(\omega)$ using a scheme similar to that of the previous section(see \SMcite{subsec:sup:numtau}). The iterative process is terminated when we have converged to a solution for $\rho(\omega)$ with sufficient accuracy.
\section{Finite temperature analysis}\mylabel{sec:sup:finiteTana}
The saddle-point equations given by \eqn{eq:sup:SPeqtau12}(a) and (b)  do not possess time reparametrization symmetry, however we expect this symmetry to gradually emerge as $\gamma\rightarrow 1$, since in this limit the system approaches the 0-dimensional SYK model, indicated by the fact $\Delta\rightarrow 1/2q$, i.e., the scaling dimension in the 0-dimensional  SYK model. Therefore assuming the reparameterization symmetry to approximately hold, we find the finite temperature solutions for $G$ and $\Sigma$ in the non-Fermi liquid regime by mapping $\tau=f(\sigma)$ such that the bilocal fields transform as
\begin{align*}
F(\tau_{1},\tau_{2})\to\tilde{F}(\tau_{1},\tau_{2})= & [f'(\tau_{1})f'(\tau_{2})]^{\Delta_F}F(f(\tau_{1}),f(\tau_{2}))
\end{align*}
where $\Delta_F$ denotes $\Delta$($\Delta_{\Sigma}$) when $F=G(\Sigma)$.
We map the line $-\infty<\tau<\infty$ at $T=0$
to $0<\tau<\beta$ using $\tau\to(\beta/\pi)\tan(\pi\tau/\beta)$
and $f'(\tau)=\partial f/\partial\tau=\sec^{2}(\pi\tau/\beta)$. This gives a scaling form for the imaginary time functions 
\begin{align*}
F(\tau)= & -A(\beta J)^{-2\Delta_F}f(\tau/\beta)\hspace{1em}\hspace{1em}\tau>0,
\end{align*}
where 
\begin{align}\mylabel{eq:sup:As}
A= &
\begin{cases} C\pi^{2\Delta-1}J^{2\Delta}\sin(\pi\Delta)\Gamma(2\Delta)& \textup{for }F=G
\\
J^{2}\frac{C^{2q-1}}{\pi^{2q-1-2\Delta_{s}}}\Gamma^{2q-1}(2\Delta)\sin^{2q-1}(\pi\Delta)J^{2\Delta_{s}}\equiv A_{\Sigma}
&\textup{for }
F=\Sigma
\end{cases}
\end{align}
and $f(x)= [\sin(\pi x)]^{-2\Delta_F}$.
From the above scaling forms for $G$ and $\Sigma$, we can obtain
the scaling forms for corresponding spectral densities, namely
\begin{align}\label{eq:sup:T_SpectralFn}
\rho_F(\omega)= & \frac{A}{J}\left(\frac{T}{J}\right)^{2\Delta_F-1}\vartheta(\omega/T)\\
\vartheta(x)= & \frac{2^{2\Delta_F-1}}{\pi^{2}}\cosh(x/2)\frac{\Gamma(\Delta_F+\ci x/2\pi)\Gamma(\Delta_F-\ci x/2\pi)}{\Gamma(2\Delta_F)}\nonumber, 
\end{align}
where $\rho_F=\rho$($\rho_\Sigma$) for $F=G$($\Sigma$). Similarly, the finite-temperature retarded functions are
\begin{align}\mylabel{eq:sup:finT_scalf}
F(\omega)= & \frac{A}{J}\left(\frac{T}{J}\right)^{2\Delta_F-1}f_F(\omega/T)
\\
f_F(x)= & -\ci\frac{2^{2\Delta_F-1}\sin(\pi\Delta_F+\ci x/2)}{\pi\sin(\pi\Delta_F)}\frac{\Gamma(\Delta_F-\ci x/2\pi)\Gamma(\Delta_F+\ci x/2\pi)}{\Gamma(2\Delta_F)}.\nonumber 
\end{align}
Unlike in the case of 0-dimensional SYK model, the above scaling forms for $G(\omega)$ and $\Sigma(\omega)$ only approximately satisfy the low-energy saddle-point equation (9) (main text) at the NFL fixed point, suggesting that the reparameterization symmetry is  not exact even at arbitrary low energies. However, as demosntrated by our results in the main text,  the reparametrization symmetry is only weakly broken at the NFL fixed point. An estimate of the degree of symmetry breaking can be obtained by analytically continuing $G(\omega),\Sigma(\omega)$ obtained from \eqn{eq:sup:finT_scalf} to $G(\iwn),\Sigma(\iwn)$ and then substituting the result in the low-energy saddle-point equation,
\begin{align}
G(\ci\omega_{n})\Sigma(\ci\omega_{n})^{\gamma}= & g_{0}\pi\csc(\gamma\pi)(e^{-\ci\gamma\pi}-1),
\end{align}
to obtain
\bea
\underbrace{\left(\frac{\Gamma(\Delta+\frac{\beta\wn}{2\pi})}{\Gamma(1-\Delta+\frac{\beta\wn}{2\pi})}\right)\left(\frac{\Gamma(\Delta_{\Sigma}+\frac{\beta\wn}{2\pi})}{\Gamma(1-\Delta_{\Sigma}+\frac{\beta\wn}{2\pi})}\right)^{\gamma}}_{\LHS(\iwn)}
\approx h(\beta,\gamma,q),
\eea
where $h$ is a $\iwn$ independent constant. Due to the absence of reparameterization invariance, $\LHS$ weakly depends on $\iwn$. However, we have checked that $\LHS(\iwn)$ is a monotonic function of $\iwn$ that rapidly converges to a constant value as $\iwn$ increases. Therefore an estimate for the amount of reparametrization symmetry breaking can be defined as
\begin{align}
\delta_{\textup{break}}(\gamma,q)=\left[\frac{\LHS(\gamma,q,\wn\rightarrow\infty)}{\LHS(\gamma,q,\wn\rightarrow0)}\right]-1
\end{align}
which measures the deviation of $\LHS$ from being a constant, such that  for $\delta_{\textup{break}}=0$ we have exact reparametrization symmetry. Using the asymptotic form, $\ln\Gamma(z)\approx(z-\frac{1}{2})\ln z-z+\frac{1}{2}\ln(2\pi)$ for $|z|\to\infty$ we get
\begin{align}
\ln\LHS(\gamma,q,\wn\rightarrow\infty)	=&	(1+\gamma)\ln\pi\notag\\
\ln\LHS(\gamma,q,\wn\rightarrow0)=&(1+\gamma)\ln\pi-\frac{(q-1)\gamma}{(1+(2q-1)\gamma)}\ln\left[\frac{\Delta(1-\Delta)}{\frac{1}{4}-4\left(\frac{1-q}{1+\gamma}\right)^{2}\Delta^{2}}\right].
\end{align}
 Hence, the error becomes
 \bea
 \delta_{\textup{break}}(\gamma,q)=\left[\frac{\Delta(1-\Delta)}{\frac{1}{4}-4\left(\frac{1-q}{1+\gamma}\right)^{2}\Delta^{2}}\right]^\frac{(q-1)\gamma}{(1+(2q-1)\gamma)}-1,
 \eea
 which we plot in \fig{fig:sup:figure5}, for $q=2,3,4$ and $\gamma_c<\gamma<1$. From the figure we see that the error becomes large near $\gamma\approx\gamma_c$ but $\delta_\mathrm{break}\to 0$ as $\gamma\rightarrow 1$. However, note that that this is only a rough estimate to demonstrate that the solutions constructed assuming reparametrization symmetry are quite close to the finite-temperature solution of infrared saddle-point equation (9) (main text) .
  \ifdefined\showfigures
\begin{figure}
\includegraphics[scale=1.25]{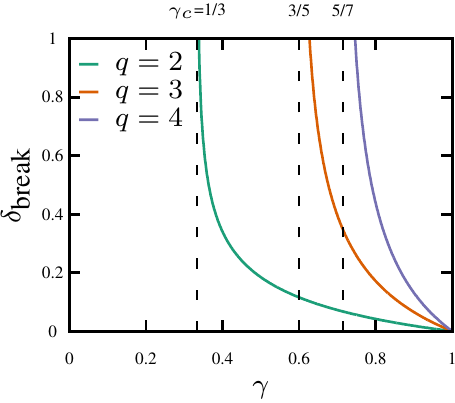}
\caption{{\bf Reparametrization symmetry breaking estimate:} Plot of $\delta_{\textup{break}}$ as a function of $\gamma$ for various $q$ giving an estimate for the error encountered when solutions constructed using  reparametrization symmetry are used to solve the low-energy saddle-point equations (9),(6b) [main text] at finite tempearture. For all $q\geq 2$ the error tends to \emph{zero} as $\gamma$ approaches $1$ and increases drastically near $\gamma_c(q)$.}
\mylabel{fig:sup:figure5}
\end{figure}
\fi \section{Thermodynamics}\mylabel{sec:sup:thermo}
The free-energy density for the system can be obtained by evaluating the action [eqn.~(5), main text] using the saddle point solutions, i.e.,
\begin{align}\mylabel{eqn:sup:freee}
F=&-T\sum\limits_{\iwn}\int\de g(\varepsilon)\ln\left[-\iwn+\varepsilon+\Sigma(\iwn)\right]-\frac{J^2}{2q}\int_0^\beta\dtau G^q(\beta-\tau)G^q(\tau)-T\sum\limits_{\iwn}\Sigma(\iwn) G(\iwn),
\end{align}
The entropy $S$ is evaluated via $S=-\partial F/\partial T$.
\paragraph{\underline{Case -- Non-Fermi liquid limit:}}\mylabel{para:sup:SNFLlim} We first derive an expression for entropy for the non-Fermi liquid case when $\gamma>\gamma_c(q)$ using the scaling solutions \eqn{eq:sup:T_SpectralFn}. From eqn.~(6b) in \eqn{eqn:sup:freee} we get
\begin{align}\mylabel{eq:sup:freeen}
F= & -\left(\frac{2q-1}{2q}\right)\underbrace{T\sum_{n}\Sigma(\ci\omega_{n})G(\ci\omega_{n})}_{F_1}
\underbrace{-T\sum\limits_{\iwn}\int\de g(\varepsilon)\ln\left[-\iwn+\varepsilon+\Sigma(\iwn)\right]}_{F_2},
\end{align}
We consider the terms $F_1$ and $F_2$ separately below and obtain their leading
low-temperature behaviors. 

 It can be shown that
\begin{align*}
\sum_{\iwn}\Sigma(\ci\omega_{n})G(\ci\omega_{n})=  \sum_{\iwn}\Sigma(\ci\omega_{n})\left(\frac{1}{N_L}\sum_{b,\mathbf{k}}\Gkwn\right)=& \frac{1}{N_L}\sum_{b,\mathbf{k}}\int d\omega\frac{(\omega-\ek)\Abkw}{1+e^{\beta\omega}},
\end{align*}
where $\Abkw=-(1/\pi)\Im\Gkwn$. Using the above we get
\begin{align}\mylabel{eq:sup:f1}
F_{1}(T)
= \frac{1}{N_L}\sum_{\mathbf{k},b}\int_{-\infty}^{\infty}d\omega\frac{(\omega-\ek)\Abkw}{1+e^{\beta\omega}}
= 
\underbrace{\int_{-\infty}^{\infty}d\omega\frac{\omega\rho(\omega)}{1+e^{\beta\omega}}}_{F_{1a}}
\underbrace{-\int_{-\infty}^{\infty}d\omega\int_{-\Lambda}^{\Lambda}\de\ \varepsilon g(\varepsilon)\frac{\rho(\varepsilon,\omega)}{1+e^{\beta\omega}}}_{F_{1b}},
\end{align}
where $\rho(\varepsilon,\omega)=-(1/\pi)\Im[(\omega-\varepsilon-\Sigma(\omega))^{-1}]$ and $\rho(\omega)=\int_{-\Lambda}^{\Lambda}d\epsilon g(\varepsilon)\rho(\varepsilon,\omega)$. Due to particle-hole symmetry, 
\begin{eqnarray}\mylabel{eq:sup:PHsymmrel}
g(-\varepsilon)=  g(\varepsilon), &
\ \ 
\Sigma'(-\omega)=  -\Sigma'(\omega),&
\ \ 
\Sigma''(-\omega)= \Sigma''(\omega),\notag\\
\rho(\varepsilon,\omega)= \rho(-\varepsilon,-\omega),&
\rho(\omega)=  \rho(-\omega).
\end{eqnarray}
where $\Sigma'$ ($\Sigma''$) are the real (imaginary) part of $\Sigma(\omega)$. Using these properties we obtain 
\begin{align*}
F_{1a}(T)= & 2\int_{0}^{\infty}d\omega\frac{\omega\rho(\omega)}{1+e^{\beta\omega}}-\int_{0}^{\infty}d\omega\ \omega\rho(\omega).
\end{align*}
Since we are going to take a temperature derivative to obtain entropy,
we subtract from above $F_{1a}(T=0)=-\int_{0}^{\infty}d\omega\ \omega\rho(\omega,T=0)$.
This takes care of the ultraviolet contribution that is not captured by
the scaling solution of  \eqn{eq:sup:T_SpectralFn}. Hence,
\begin{align*}
F_{1a}(T)-F_{1a}(T=0)= & -\int_{0}^{\infty}d\omega\omega[\rho(\omega)-\rho(\omega,T=0)]+2\int_{0}^{\infty}d\omega\frac{\omega\rho(\omega)}{1+e^{\omega/T}}.
\end{align*}
Using the scaling form of \eqn{eq:sup:T_SpectralFn}, we obtain for the two integrals above
\begin{align*}
2\int_{0}^{\infty}d\omega\frac{\omega\rho(\omega)}{1+e^{\omega/T}}= & 2\frac{A}{J^{2\Delta}}T^{2\Delta+1}\int_{0}^{\infty}du\frac{u\vartheta(u)}{1+e^{u}}\\
\int_{0}^{\infty}d\omega\omega[\rho(\omega)-\rho(\omega,T=0)]= & \frac{A}{J^{2\Delta}}T^{2\Delta+1}\int_{0}^{\infty}duu\left[\vartheta(u)-c_{1}u^{2\Delta-1}\right].
\end{align*}
Evidently, the integral in the first line is convergent, so there
is no need for an ultraviolet cutoff. To check whether the integral in
the second line is convergent, we check how the integrand behaves
as $u\to\infty$. It can be shown that $\vartheta(u)=c_{1}u^{2\Delta-1}+c_{2}u^{2\Delta-3}+\dots$ for $u\to\infty$ (see \mycite{Parcollet1998}).
This gives an integrand $c_{2}u^{2\Delta-2}$. For $\gamma_{c}\leq\gamma\leq1$ we have 
$-2+1/q\leq2\Delta-2\leq-2+2/(2q-1)$. Hence for $q\geq2$, the above
integral is always convergent. From the above we then find $F_{1a}\sim T^{2\Delta+1}$
so the contribution to entropy goes as
\begin{align}\mylabel{eq:sup:s1a}
S_{1a}\sim & T^{2\Delta}.
\end{align}
Now, for $F_{1b}$ in \eqn{eq:sup:f1}, using particle-hole symmetry we get
\begin{align}
F_{1b}(T)= & \int_{0}^{\infty}d\omega\int_{-\Lambda}^{\Lambda}\de\ \varepsilon g(\varepsilon)\rho(\varepsilon,\omega)= -\frac{1}{\pi}\Im\left[\int_{0}^{\infty}d\omega\int_{-\Lambda}^{\Lambda}\de\ \varepsilon g(\varepsilon)G(\varepsilon,\omega)\right],
\end{align}
 where $G(\varepsilon,\omega)\equiv(\omega-\varepsilon-\Sigma(\omega))^{-1}$. We can perform the integral over $\varepsilon$, as in \eqn{eq:sup:Gzbeta}, in terms of the incomplete beta function and expand around $\Lambda\rightarrow\infty$ for $\omega\ll \Sigma(\omega)$ to get
\begin{align*}
\int_{-\Lambda}^{\Lambda}\de\ \varepsilon g(\varepsilon)G(\varepsilon,\omega)= & -g_{0}\left[\pi e^{-\ci(\pi/2)(\gamma+1)}\Sigma(\omega)^{1-\gamma}\sec(\gamma\pi/2)+\frac{2}{\gamma}\Lambda^{1-\gamma}+\dots\right].
\end{align*}
As earlier, we subtract $F_{1b}(T=0)$ from $F_{1b}(T)$ to obtain
\begin{align*}
F_{1b}(T)-F_{1b}(T=0)= & \frac{1}{\pi}\Im\left[g_{0}\pi\frac{e^{-\ci(\pi/2)(\gamma+1)}}{\cos(\gamma\pi/2)}\int_{0}^{\infty}d\omega\left(\Sigma(\omega)^{1-\gamma}-\Sigma(\omega,T=0)^{1-\gamma}\right)\right]
\end{align*}
Again substituting the scaling form for $\Sigma(\omega)$ from \eqn{eq:sup:finT_scalf}, 
\begin{align}\mylabel{eq:sup:f1bTmf1b0}
F_{1b}(T)-F_{1b}(T=0)= & \frac{1}{\pi}\Im\left[g_{0}\pi\frac{e^{-\ci(\pi/2)(\gamma+1)}}{\cos(\gamma\pi/2)}\left(\frac{A_{\Sigma}}{J^{2\Delta_{\Sigma}}}\right)^{1-\gamma}T^{(2\Delta_{\Sigma}-1)(1-\gamma)+1}\int_{0}^{\infty}du\left(f_{\Sigma}(u)^{1-\gamma}-c_{1}^{1-\gamma}u^{(2\Delta_{\Sigma}-1)(1-\gamma)}\right)\right].
\end{align}
Using the the series expansion $f_{\Sigma}(u)=c_{1}u^{2\Delta_{\Sigma}-1}+c_{2}u^{2\Delta_{\Sigma}-3}+\dots$
for $u\to\infty$, we get
\begin{align}
f_{\Sigma}(u)^{1-\gamma}-c_{1}^{1-\gamma}u^{(2\Delta_{\Sigma}-1)(1-\gamma)}\simeq & (1-\gamma)c_{1}^{-\gamma}c_{2}u^{(2\Delta_{\Sigma}-1)(1-\gamma)-2}.
\end{align}
Since $\Delta_{\Sigma}\leq1$ for $\gamma_{c}\leq\gamma\leq1$, the integrand in \eqn{eq:sup:f1bTmf1b0}
is convergent for $u\to\infty$. Hence we find $F_{1b}\sim T^{(2\Delta_{\Sigma}-1)(1-\gamma)+1}$ implying that its contribution to the low-temperature entropy goes as
\begin{align}\label{eq:sup:s1b}
S_{1b}\sim & T^{(2\Delta_{\Sigma}-1)(1-\gamma)}.
\end{align}
Comparing this with $S_{1a}$ in \eqn{eq:sup:s1a} we find that $S_{1b}$ dominates over $S_{1a}$ for $\gamma_{c}\leq\gamma\leq1$ as $T\to 0$. Also, it can be shown from \eqn{eq:sup:f1bTmf1b0} that $S_{1b}\to 0$ in the limit $\gamma\to 1$, i.e., approaching the 0-dimensional SYK limit.

We now evaluate the contribution to entropy from the $F_2$ in \eqn{eq:sup:freeen}. Following Ref.\onlinecite{Parcollet1998}, it can be shown that
\begin{align}\label{eq:sup:f2}
F_{2}= & \int \de\ g(\varepsilon)\int_{-\infty}^{\infty}\frac{d\omega}{\pi}\left(\arctan\left(\frac{G'(\varepsilon,\omega)}{G''(\varepsilon,\omega)}\right)-\frac{\pi}{2}\right)n_{F}(\omega)=\int_{-\infty}^{\infty}\frac{d\omega}{\pi}n_{F}(\omega)\int \de\  g(\varepsilon)\left[\arctan\left(\frac{\omega-\epsilon-\Sigma'}{|\Sigma''|}\right)-\frac{\pi}{2}\right],
\end{align}
where $G=G'+\ci G''$, in terms of real and imaginary parts, and $n_F(\omega)=1/(e^{\beta\omega}+1)$ is the Fermi function. 
To evaluate the above, we use the identity
\begin{align*}
\arctan\left(\frac{\omega-\varepsilon-\Sigma'}{|\Sigma''|}\right)-\frac{\pi}{2}= & \Re\left[|\Sigma''|\int_{\infty}^{1}\frac{dx}{x^{2}}\frac{1}{(\omega-\varepsilon-\Sigma')-\ci\Sigma''/x}\right].
\end{align*}
The integral over $\varepsilon$ in \eqn{eq:sup:f2} can now be performed, as in \eqn{eq:sup:Gz}, giving us
\begin{align*}
F_{2}= & \Re\left[g_{0}\pi\csc(\gamma\pi)(1-e^{\ci\gamma\pi})\int_{-\infty}^{\infty}\frac{d\omega}{\pi}n_{F}(\omega)\frac{|\Sigma''(\omega)|}{(\omega-\Sigma'(\omega))^{\gamma}}\int_{\infty}^{1}\frac{dx}{x^{2-\gamma}}\frac{1}{\left(x-\ci\frac{\Sigma''}{\omega-\Sigma''}\right)^{\gamma}}\right].
\end{align*}
The integral over $x$ can also be easily performed to obtain
\begin{align*}
F_{2}= & \Re\left[\ci\frac{g_{0}\pi}{1-\gamma}\csc(\gamma\pi)(1-e^{\ci\gamma\pi})\int_{-\infty}^{\infty}\frac{d\omega}{\pi}n_{F}(\omega)\left(\left(\omega-\Sigma'(\omega)-\ci\Sigma''(\omega)\right)^{1-\gamma}-(\omega-\Sigma'(\omega))^{1-\gamma}\right)\right].
\end{align*}
Now we use the finite temperature scaling form of $\Sigma(\omega)$ from
\eqn{eq:sup:finT_scalf} and neglect $\omega\ll\Sigma(\omega)$ in the above integral. In particular we have,
\begin{align*}
\begin{array}{ll}
\Sigma'(\omega)=  \tilde{A}_{\Sigma}T^{2\Delta_{\Sigma}-1}\cot(\pi\Delta_{\Sigma})\sigma'\left(\frac{\omega}{T}\right)
&
\Sigma''(\omega)=  -\tilde{A}_{\Sigma}T^{2\Delta_{\Sigma}-1}\sigma''\left(\frac{\omega}{T}\right)\\
\sigma'(x)=  \sinh\left(\frac{x}{2}\right)B\left(\Delta_{\Sigma}+\ci\frac{x}{2\pi},\Delta_{\Sigma}-\ci\frac{x}{2\pi}\right)
&
\sigma''(x)=  \cosh\left(\frac{x}{2}\right)B\left(\Delta_{\Sigma}+\ci\frac{x}{2\pi},\Delta_{\Sigma}-\ci\frac{x}{2\pi}\right)
\end{array}
\end{align*}
where $B(x,y)=\Gamma(x)\Gamma(y)/\Gamma(x+y)$ is the complete beta function,
$\tilde{A_{\Sigma}}=A_{\Sigma}/\pi J^{2\Delta_{\Sigma}}$ with $A_{\Sigma}$ is given in
\eqn{eq:sup:As} and $1-1/2q\leq\Delta_{\Sigma}\leq1$. This results in 
\begin{align}\mylabel{eq:sup:f2-1}
F_{2}= & \tilde{A}_{\Sigma}^{1-\gamma}T^{(2\Delta_{\Sigma}-1)(1-\gamma)+1}\Re\left[\ci(-1)^{1-\gamma}\frac{g_{0}\pi}{1-\gamma}\csc(\gamma\pi)(1-e^{\ci\gamma\pi})\right.\nonumber \\
 &\left. \times\int_{-\infty}^{\infty}\frac{du}{\pi}\frac{1}{e^{u}+1}\sigma''(u)^{1-\gamma}\left(\left(\cot(\pi\Delta_{\Sigma})\tanh(u/2)-\ci\right)^{1-\gamma}
 -(\cot(\pi\Delta_{\Sigma})\tanh(u/2))^{1-\gamma}\right)\right].
\end{align}
The integral over $u$ is convergent since $
B(x+\ci y,x-\ci y)= |\Gamma(x+\ci y)|^{2}/\Gamma(2x)$ and
$|\Gamma(x+\ci y)|\overset{y\to\infty}{=} \sqrt{2\pi}e^{-|y|/2}|y|^{x-1/2}$ implying
$\sigma''(u)\to e^{-|u|/2}$ as $|u|\to\infty$. Hence we find $F_{2}\sim T^{(2\Delta_{\Sigma}-1)(1-\gamma)+1}$ with the corresponding contribution to entropy being
\begin{align}
S_{2}\sim & T^{(2\Delta_{\Sigma}-1)(1-\gamma)}\mylabel{eq:sup:s2}.
\end{align}
 $S_{2}$ has the same temperature dependence as $S_{1b}$ (see \eqn{eq:sup:s1b}) and $S_{1b}+S_{2}$ dominates the low-temperature entropy for $\gamma_{c}\leq\gamma\leq1$, so that the total entropy  
\begin{align*}
S\sim & T^{(2\Delta_{\Sigma}-1)(1-\gamma)}.
\end{align*}
The above implies that the entropy increases rapidly, much faster than $T$, with increasing temperature as $\gamma$ approaches $1$. 

The residual entropy of the 0-dimensional SYK model can be recovered by carefully taking the limit $\gamma\to1$ in \eqn{eq:sup:f2-1}. In this limit,  $g_{0}/\csc(\gamma\pi)\to1/2\pi$. Using the identity, $\ln x=\lim_{x\to0}(x^{n}-1)/n$
\begin{align*}
S(T=0)= & -\int_{-\infty}^{\infty}\frac{du}{\pi}\frac{1}{e^{u}+1}\left(\arctan(\cot(\pi\Delta_{\Sigma})\tanh(u/2))-\frac{\pi}{2}\right).
\end{align*}
For $\gamma=1$ we have $\Delta=1/2q$, $\Delta_{\Sigma}=1-1/2q$ and $\cot(\pi\Delta_{\Sigma})=-\cot(\pi\Delta)$.
Hence,
\begin{align}\mylabel{eq:sup:SlimitSYK}
S(\gamma\to1,T=0)= & \int_{-\infty}^{\infty}\frac{du}{\pi}\frac{1}{e^{u}+1}\left(\arctan(\cot(\pi\Delta)\tanh(u/2))+\frac{\pi}{2}\right)
\end{align}
which is exactly the expression
for zero-temperature entropy for the 0-dimensional SYK model.
\paragraph{\underline{Case -- Fermi liquid limit:}}\mylabel{para:sup:SFLlim} 
The free-energy in the LFL regime can be obtained by neglecting the self-energy contributions in \eqn{eq:sup:freeen} giving us 
\begin{align}
F= & 
-T\sum\limits_{\iwn}\int\de g(\varepsilon)\ln\left[-\iwn+\varepsilon\right].
\end{align}
The sum over the Matsubara frequencies can be  easily evaluated  to get
\begin{align}
F=
-T\int\de g(\varepsilon)\ln\left[1+\exp(\varepsilon/T)\right]=&-g_0\int_{0}^{\Lambda}\de\  \varepsilon^{1-\gamma}
-2g_0T\int_{0}^{\Lambda}\de\  \varepsilon^{-\gamma}
\ln\left[1+\exp(-\varepsilon/T)\right].
\end{align}
We perform a change of variables $\varepsilon/T\to\varepsilon$ in the second-integral appearing above followed by a integration by parts to obtain
\begin{align}\mylabel{eq:sup:freeenFL}
F=&-g_0\int_{0}^{\Lambda}\de\  \varepsilon^{1-\gamma}
-T\ln(1+e^{-\Lambda/T})
-T^{2-\gamma}\Lambda^{\gamma-1}\int_{0}^{\Lambda/T\rightarrow\infty}\de\  \frac{\varepsilon^{1-\gamma}}{1+e^{\tilde{\varepsilon}}}
,
\end{align}
where we have taken the limit $\Lambda/T\rightarrow\infty$ in the second integral since we are interested in low-temperature behavior. Clearly all the integrals appearing above are convergent, hence the dominant contribution to the free-energy comes from the last term in \eqn{eq:sup:freeenFL} implying $F\sim T^{2-\gamma}$ and
\bea
S\sim T^{1-\gamma}.
\eea
\section{Lyapunov exponent calculation}\mylabel{sec:sup:lyapexp}
The 0-dimensional SYK model saturates the upper bound of chaos \mycite{KitaevKITP}. The chaos in this case is characterized via intermediate-time behavior of out-of-time-ordered (OTO) correlation function. To this end, we compute the following OTO correlators for our lattice mode withl $q=2$, namely
\begin{align}
\FOne_{\alpha\alpha'}(\vtr{x}-\vtr{x'};t,t)=&\frac{1}{N^2}\sum_{ij}\langle \cd_{i\alpha \vtr{x}}(t)y \cd_{j\alpha' \vtr{x}'}(0)y c_{i\alpha \vtr{x}}(t) y c_{j\alpha' \vtr{x'}}(0)y \rangle
\notag\\
\FTwo_{\alpha\alpha'}(\vtr{x}-\vtr{x'};t,t)=&\frac{1}{N^2}\sum_{ij}\langle c_{i\alpha \vtr{x}}(t)y \cd_{j\alpha' \vtr{x'}}(0) y\cd_{i\alpha \vtr{x}}(t)y c_{j\alpha' \vtr{x'}}(0)y \rangle
\end{align}
where $y^4=e^{-\beta \cH}/Z$ , $Z$ is the partition function and $\langle\cdots\rangle$ denotes a trace followed by disorder averaging. The onset of chaos is determined by the $O(1/N)$ terms that appear in the large-$N$ expansion of the above correlators. Over an intermediate time window, $\lambda_L^{-1}\ll t<\lambda_L^{-1}\log{N}$, these OTO correlators are expected to decay as $f_0-f_1/N e^{\lambda_L t}+O(1/N^2)$ where $\lambda_L$ is the Lyapunov exponent. Following Refs.\cite{KitaevKITP,Maldacena2016,BanerjeeAltman2016}, the $O(1/N)$ parts (see \fig{fig:sup:figure6}) of the OTO correlators $F_{1,2}$ can be obtained via the following self-consistent equations 
\begin{align}\mylabel{eq:sup:OTOCSCeq}
\FOne_{\alpha\alpha'}(\vtr{x}-\vtr{x'};t_{1,2})=&
\sum_{\gamma,\vtr{x_1}}\int_{-\infty}^{+\infty}\dt_3\dt_4\left[\ 
\KOO_{\alpha\gamma}(\vtr{x}-\vtr{x_1};t_{1,2,3,4})\FOne_{\gamma\alpha'}(\vtr{x_1}-\vtr{x'};t_{3,4})
+
\KOT_{\alpha\gamma}(\vtr{x}-\vtr{x_1};t_{1,2,3,4})\FTwo_{\gamma\alpha'}(\vtr{x_1}-\vtr{x'};t_{3,4})\right]\notag\\
\FTwo_{\alpha\alpha'}(\vtr{x}-\vtr{x'};t_{1,2})=&
\sum_{\gamma,\vtr{x_1}}\int_{-\infty}^{+\infty}\dt_3\dt_4\left[\ 
\KOT_{\alpha\gamma}(\vtr{x}-\vtr{x_1};t_{1,2,3,4})\FOne_{\gamma\alpha'}(\vtr{x_1}-\vtr{x'};t_{3,4})
+
\KTT_{\alpha\gamma}(\vtr{x}-\vtr{x_1};t_{1,2,3,4})\FTwo_{\gamma\alpha'}(\vtr{x_1}-\vtr{x'};t_{3,4})\right].
\end{align}
Here $t_{1,2}\equiv t_1,t_2$ and $t_{1,2,3,4}\equiv t_1,t_2,t_3,t_4$.
\ifdefined\showfigures
\begin{figure}
\includegraphics[scale=1.0]{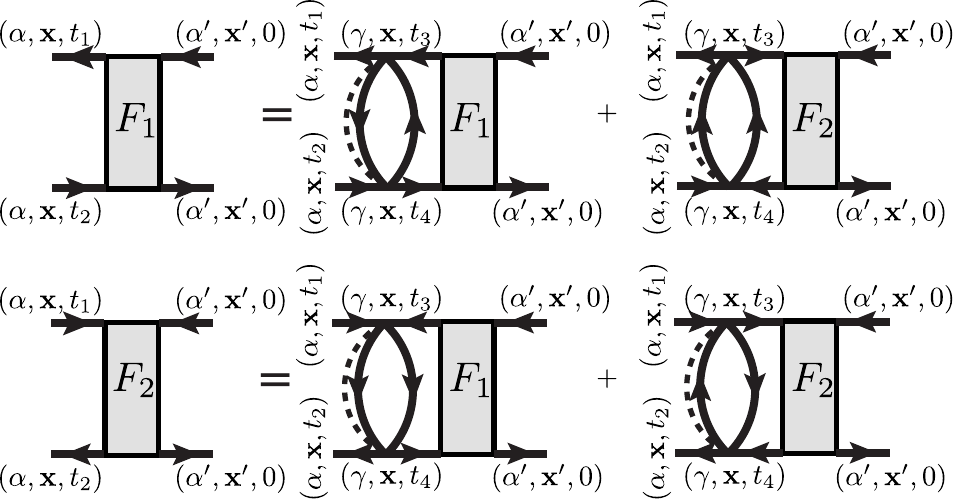}
\caption{Diagrammatic representation of the self-consistent equations for the $1/N$ parts of the out-of-time-order functions. Here $\alpha,\alpha'$ are the fermion color indices, $\vtr{x},\vtr{x'}$ are lattice coordinates  and solid lines represent various Green's functions.}
\mylabel{fig:sup:figure6}
\end{figure}
\fi 
The kernel $K$s are given by
\begin{align}\mylabel{eq:sup:Ks}
\KOO_{\alpha\gamma}(\vtr{x}-\vtr{x_1};t_{1,2,3,4})=&\ 2J^2
G^A_{\gamma\alpha}(\vtr{x_1}-\vtr{x};t_{31})
G^R_{\alpha\gamma}(\vtr{x}-\vtr{x_1};t_{24})
G^+_{lr}(t_{43})
G^-_{lr}(t_{34})
\notag\\
\KOT_{\alpha\gamma}(\vtr{x}-\vtr{x_1};t_{1,2,3,4})=&-J^2
G^A_{\gamma\alpha}(\vtr{x_1}-\vtr{x};t_{31})
G^R_{\alpha\gamma}(\vtr{x}-\vtr{x_1};t_{24})
G^+_{lr}(t_{43})
G^+_{lr}(t_{43})
\notag\\
\KTO_{\alpha\gamma}(\vtr{x}-\vtr{x_1};t_{1,2,3,4})=&-J^2
G^A_{\gamma\alpha}(\vtr{x_1}-\vtr{x};t_{42})
G^R_{\alpha\gamma}(\vtr{x}-\vtr{x_1};t_{13})
G^-_{lr}(t_{34})
G^-_{lr}(t_{34})
\notag\\
\KTT_{\alpha\gamma}(\vtr{x}-\vtr{x_1};t_{1,2,3,4})=&\ 2J^2
G^A_{\gamma\alpha}(\vtr{x_1}-\vtr{x};t_{42})
G^R_{\alpha\gamma}(\vtr{x}-\vtr{x_1};t_{13})
G^+_{lr}(t_{43})
G^-_{lr}(t_{34})
\end{align}
where $t_{ij}=t_i-t_j$, $G^A$($G^R$) is the advanced(retarded) Green's functions and $G_{lr}^\pm=G(\ci t\pm\beta/2)$ are Wightmann correlators obtained by contracting fermions with two different Keldysh indices (see \mycite{Maldacena2016,BanerjeeAltman2016} for details). The convolution over $\vtr{x_1}$ in \eqn{eq:sup:OTOCSCeq} can be made into a product by moving to momentum representation, resulting in
\begin{align}\mylabel{eq:sup:OTOCSCeq2}
\FOne_{\alpha\alpha'}(\vtr{q};t_{1,2})=&
\sum_{\gamma}\int_{-\infty}^{+\infty}\dt_3\dt_4\left[\ 
\KOO_{\alpha\gamma}(\vtr{q};t_{1,2,3,4})\FOne_{\gamma\alpha'}(\vtr{q};t_{3,4})
+
\KOT_{\alpha\gamma}(\vtr{q};t_{1,2,3,4})\FTwo_{\gamma\alpha'}(\vtr{q};t_{3,4})\right]\notag\\
\FTwo_{\alpha\alpha'}(\vtr{q};t_{1,2})=&
\sum_{\gamma}\int_{-\infty}^{+\infty}\dt_3\dt_4\left[\ 
\KOT_{\alpha\gamma}(\vtr{q};t_{1,2,3,4})\FOne_{\gamma\alpha'}(\vtr{q};t_{3,4})
+
\KTT_{\alpha\gamma}(\vtr{q};t_{1,2,3,4})\FTwo_{\gamma\alpha'}(\vtr{q};t_{3,4})\right],
\end{align}
where $K(\vtr{q})=\sum\limits_\vtr{x} K(\vtr{x}) e^{-\ci \vtr{q}\cdot\vtr{x}}$ and $F(\vtr{q})=\sum\limits_\vtr{x} F(\vtr{x}) e^{-\ci \vtr{q}\cdot\vtr{x}}$, respectively. To evaluate the Lyapunov exponent $\lambda_L$ we use the following chaos ansatz \mycite{KitaevKITP,Maldacena2016}
\bea
F^{(i)}_{\alpha\alpha'}(\vtr{q},t_{1,2})=e^{\lambda_L(\vtr{q})(t_1+t_2)/2}f^{(i)}_{\alpha\alpha'}(\vtr{q},t_{12})~~~i=1,2.
\eea
 Following Ref. \onlinecite{BanerjeeAltman2016}, the time integrals in \eqn{eq:sup:OTOCSCeq2} can be performed and the final result can be written as an eigenvalue equation in the real-frequency domain 
\begin{equation}\mylabel{eq:sup:fqw}
|f(\vtr{q},\omega)\rangle=
\int_{-\infty}^{+\infty}\dw'\frac{1}{2\pi}
\left[
\begin{array}{cc}
2J^{2}g_{1}(\omega-\omega')
\left[
\begin{array}{cc}
\tilde{K}_{RR}& \tilde{K}_{RB}\\
\tilde{K}_{BR} & \tilde{K}_{BB}
\end{array}
\right]_{(\vtr{q},-\omega,\lambda_L)}
 & -J^{2}g_{2}(\omega'-\omega)
\left[
\begin{array}{cc}
\tilde{K}_{RR}& \tilde{K}_{RB}\\
\tilde{K}_{BR} & \tilde{K}_{BB}
\end{array}
\right]_{(\vtr{q},-\omega,\lambda_L)}
\\
-J^{2}g_{2}(\omega-\omega')
\left[
\begin{array}{cc}
\tilde{K}_{RR}& \tilde{K}_{RB}\\
\tilde{K}_{BR} & \tilde{K}_{BB}
\end{array}
\right]_{(\vtr{q},\omega,\lambda_L)}
 & 2J^{2}g_{1}(\omega-\omega')
\left[
\begin{array}{cc}
\tilde{K}_{RR}& \tilde{K}_{RB}\\
\tilde{K}_{BR} & \tilde{K}_{BB}
\end{array}
\right]_{(\vtr{q},\omega,\lambda_L)}
\end{array}
\right]
|f(\vtr{q},\omega')\rangle,
\end{equation}
where 
\begin{equation}
|f(\vtr{q},\omega)\rangle\equiv
\left[
\begin{array}{c}
f_{RR}^{(1)}(\vtr{q},\omega)\\
f_{BR}^{(1)}(\vtr{q},\omega)\\
f_{RR}^{(2)}(\vtr{q},\omega)\\
f_{BR}^{(2)}(\vtr{q},\omega)
\end{array}
\right]
\end{equation}
and
\begin{align}\mylabel{eq:sup:GARw}
\tilde{K}_{\alpha\alpha'}(\vtr{q},\omega,\lambda_L)=\frac{1}{N_k}\sum_\vtr{k}G_{\alpha\alpha'}^{R}(q+k,\omega+\ci\lambda_L/2)G_{\alpha'\alpha}^{A}(\vtr{k},\omega-\ci\lambda_L/2).
\end{align}
We now briefly discuss about the various Green's functions that appear in \eqn{eq:sup:fqw} and \eqn{eq:sup:GARw}. The functions $g_1$ and $g_2$ are related to the Wightmann correlators in \eqn{eq:sup:Ks} as
\begin{align}
g_{1}(\omega)=&\int\dt\ G_{lr}^{+}(-t)G_{lr}^{-}(t)e^{\ci\omega t}\notag\\
g_{2}(\omega)=&\int\dt\ G_{lr}^{+}(t)^2e^{\ci\omega t}=\int\dt\ G_{lr}^{-}(t)^2e^{\ci\omega t}.
\end{align}
The Wightmann corellators $G_{lr}^{\pm}(t)$ can be found by analytically continuing $G(\tau)$ as shown below
\begin{align}
G_{lr}^{+}(t)=&\ci G((\tau>0)\rightarrow \ci t+\frac{\beta}{2})\notag\\
G_{lr}^{-}(t)=&\ci G((\tau<0)\rightarrow \ci t-\frac{\beta}{2}),
\end{align}
which can also be written in the real frequency domain by using the spectral representation for $G(\tau)$ as
\begin{align}
G_{lr}^{\pm}(\omega)=\mp\ci\frac{\pi\rho(\omega)}{\cosh(\omega\beta/2)},
\end{align}
where $\rho(\omega)=(-1/\pi)\mathrm{Im}[G(\omega^+)]$. The advanced and retarded Green's functions can be obtained by analytically continuing the following momentum-dependent Green's function
\begin{align}
G_{\alpha\alpha'}(\vk,\iwn)=&\sum_{\vtr{x}}\int_0^\beta\dtau\ G_{\alpha\alpha'}(\vtr{x},\tau)e^{\iwn\tau}e^{-\ci \vk\cdot \vtr{x}},
\end{align} 
where $G_{\alpha\alpha'}(\vtr{x}-\vtr{x'},\tau)=-\langle c_{\alpha \vtr{x}}(\tau)\cb_{\alpha' \vtr{x'}}(0)\rangle$. $G_{\alpha\alpha'}(\vk,\iwn)$ can be written in terms of the Bloch fermion Green's functions $G_{\pm}(\vk,\iwn)$, e.g., for a dispersion with even $p$ in \eqn{sup:eq:low_energy_disp} we get
\begin{align}\mylabel{eq:sup:GtoGpm}
G_{RR}(\vk,\iwn)=G_{BB}(\vk,\iwn)=&\frac{1}{2}\left(G_{-}(\vk,\iwn)+G_{+}(\vk,\iwn)\right)\notag\\
G_{RB}(\vk,\iwn)=G_{BR}(\vk,\iwn)=&
\frac{1}{2}\left(G_{-}(\vk,\iwn)-G_{+}(\vk,\iwn)\right),
\end{align}
where $G_{\pm}(\vk,\iwn)=(\iwn-\varepsilon_{\pm}(\vk)-\Sigma(\iwn))^{-1}$ [eqn.~(7), main text]. At this point we set the dimension $d$ to be one and since we are interested in the low-energy physics we use
\begin{align}
\varepsilon_{\pm}(\vk)=\pm\Lambda\left|\frac{k}{\pi}\right|^p,
\end{align}
instead of the high energy form in \eqn{eq:sup:high_energy_disp} with $\Lambda$ playing the role of the single-particle bandwidth. We mention here that such a choice of parameters is used only to make the subsequent calculations easier and does not affect the final results for our system. Using \eqn{eq:sup:GtoGpm} and the symmetries available in our system, the various $\tilde{K}$s in \eqn{eq:sup:fqw} get related to each other. For example, because of color interchange symmetry, i.e. 
$G_{RR}(\vk,\iwn)=G_{BB}(\vk,\iwn)$ and $G_{RB}(\vk,\iwn)=G_{BR}(\vk,\iwn)$, we get
\begin{align}
\tilde{K}_{RR}(\vtr{q},\omega,\lambda_L)=&\tilde{K}_{BB}(\vtr{q},\omega,\lambda_L)\equiv\tilde{K}_1(\vtr{q},\omega,\lambda_L)\notag\\
\tilde{K}_{RB}(\vtr{q},\omega,\lambda_L)=&\tilde{K}_{BR}(\vtr{q},\omega,\lambda_L)\equiv\tilde{K}_2(\vtr{q},\omega,\lambda_L).
\end{align}
Also due to band inversion symmetry, i.e. $\varepsilon_{\pm}(\vk)=\varepsilon_{\pm}(-\vk)$ and the relation $G_{-(+)}^{A}(\vk,\omega-\ci\lambda)=\left[G_{-(+)}^{R}(\vk,\omega+\ci\lambda)\right]^{*}
 $ we have
\begin{align}
\tilde{K}_{1(2)}(\vtr{q},\omega,\lambda_L)=\tilde{K}_{1(2)}(-\vtr{q},\omega,\lambda_L)=\tilde{K}_{1(2)}(\vtr{q},\omega,\lambda_L)^*.
\end{align}
Finally, due to the particle-hole symmetry
\begin{align}
\tilde{K}_{1(2)}(\vtr{q},\omega,\lambda_L)=&\tilde{K}_{1(2)}(\vtr{q},-\omega,\lambda_L)\notag\\
g_{2}(\omega)=&g_{2}(-\omega).
\end{align}
With the aid of the above relations, the eigenvalue \eqn{eq:sup:fqw} attains the following nice structure
\begin{equation}
|f(\vtr{q},\omega)\rangle=
\int_{-\infty}^{+\infty}\dw'\frac{1}{2\pi}
\left[
\begin{array}{cc}
2J^{2}g_{1}(\omega-\omega')
\left[
\begin{array}{cc}
\tilde{K}_{1}& \tilde{K}_{2}\\
\tilde{K}_{2} & \tilde{K}_{1}
\end{array}
\right]_{(\vtr{q},-\omega,\lambda_L)}
 & -J^{2}g_{2}(\omega-\omega')
\left[
\begin{array}{cc}
\tilde{K}_{1}& \tilde{K}_{2}\\
\tilde{K}_{2} & \tilde{K}_{1}
\end{array}
\right]_{(\vtr{q},-\omega,\lambda_L)}
\\
-J^{2}g_{2}(\omega-\omega')
\left[
\begin{array}{cc}
\tilde{K}_{1}& \tilde{K}_{2}\\
\tilde{K}_{2} & \tilde{K}_{1}
\end{array}
\right]_{(\vtr{q},\omega,\lambda_L)}
 & 2J^{2}g_{1}(\omega-\omega')
\left[
\begin{array}{cc}
\tilde{K}_{1}& \tilde{K}_{2}\\
\tilde{K}_{2} & \tilde{K}_{1}
\end{array}
\right]_{(\vtr{q},\omega,\lambda_L)}
\end{array}
\right]
|f(\vtr{q},\omega')\rangle,
\end{equation}
which can be simplified by transforming $f\rightarrow h$ given by
\begin{align}
\left[
\begin{array}{c}
h_{\vtr{q}}^{(1)}(\omega)\\
h_{\vtr{q}}^{(2)}(\omega)\\
h_{\vtr{q}}^{(3)}(\omega)\\
h_{\vtr{q}}^{(4)}(\omega)
\end{array}
\right]=
\left[
\begin{array}{c}
f_{RR}^{(1)}(\vtr{q},\omega)+f_{BR}^{(1)}(\vtr{q},\omega)\\
f_{RR}^{(1)}(\vtr{q},\omega)-f_{BR}^{(1)}(\vtr{q},\omega)\\
f_{RR}^{(2)}(\vtr{q},\omega)+f_{BR}^{(2)}(\vtr{q},\omega)\\
f_{RR}^{(2)}(\vtr{q},\omega)-f_{BR}^{(2)}(\vtr{q},\omega)
\end{array}
\right],
\end{align}
resulting in the following two decoupled eigenvalue equations
\begin{align}
\left[
\begin{array}{c}
h_{\vtr{q}}^{(1)}(\omega)\\
h_{\vtr{q}}^{(3)}(\omega)
\end{array}
\right]
=&\int\dw'\frac{J^{2}}{2\pi}\Kintra(\vtr{q},\omega,\lambda_L)
\left[
\begin{array}{cc}
2g_{1}(\omega-\omega') & -g_{2}(\omega-\omega')\\
-g_{2}(\omega-\omega') & 2g_{1}(\omega-\omega')
\end{array}
\right]
\left[
\begin{array}{c}
h_{\vtr{q}}^{(1)}(\omega')\mylabel{eq:sup:OTOCevaleq1}\\
h_{\vtr{q}}^{(3)}(\omega')
\end{array}
\right]\\
\left[
\begin{array}{c}
h_{\vtr{q}}^{(2)}(\omega)\\
h_{\vtr{q}}^{(4)}(\omega)
\end{array}
\right]
=&\int\dw'\frac{J^{2}}{2\pi}\Kinter(\vtr{q},\omega,\lambda_L)
\left[
\begin{array}{cc}
2g_{1}(\omega-\omega') & -g_{2}(\omega-\omega')\\
-g_{2}(\omega-\omega') & 2g_{1}(\omega-\omega')
\end{array}
\right]
\left[
\begin{array}{c}
h_{\vtr{q}}^{(2)}(\omega')\\
h_{\vtr{q}}^{(4)}(\omega')
\end{array}
\right]\mylabel{eq:sup:OTOCevaleq2}.
\end{align}
The new Kernels $\Kintra,\Kinter$ are related to the old ones by
\begin{align}
\Kintra(\vtr{q},\omega,\lambda_L)=&\tilde{K}_{1}(\vtr{q},\omega,\lambda_L)+\tilde{K}_{2}(\vtr{q},\omega,\lambda_L)\notag\\
\Kinter(\vtr{q},\omega,\lambda_L)=&\tilde{K}_{1}(\vtr{q},\omega,\lambda_L)-\tilde{K}_{2}(\vtr{q},\omega,\lambda_L).
\end{align}
The meaning of the subscripts intra/inter in the Kernels becomes evident when we write them explicitly in terms of the Bloch fermion Green's functions, i.e., 
\begin{align}
\Kintra(\vtr{q},\omega,\lambda_L)=&\frac{1}{N_{k}}\sum_{k}\frac{1}{2}\left(G_{+}^{R}(\vtr{q}+\vk,\omega+\ci\lambda_L/2)G_{+}^{A}(\vk,\omega-\ci\lambda_L/2)+G_{-}^{R}(\vtr{q}+\vk,\omega+\ci\lambda_L/2)G_{-}^{A}(\vk,\omega-\ci\lambda_L/2)\right)\notag\\
\Kinter(\vtr{q},\omega,\lambda_L)=&\frac{1}{N_{k}}\sum_{\vk}\frac{1}{2}\left(G_{+}^{R}(\vtr{q}+\vk,\omega+\ci\lambda_L/2)G_{-}^{A}(\vk,\omega-\ci\lambda_L/2)+G_{-}^{R}(\vtr{q}+\vk,\omega+\ci\lambda_L/2)G_{+}^{A}(\vk,\omega-\ci\lambda_L/2)\right),
\end{align}
which shows that the intraband mode involves the sum of terms that operate within a band whereas the interband mode involves terms that operate across bands. Henceforth we focus on the $\vtr{q}=\vtr{0}$ mode of the above equations, which can then be explicitly written as
\begin{align}\mylabel{eq:sup:Kzero_final}
\Kzintra(\omega)=&\int_{-\Lambda}^\Lambda\de\frac{g(\varepsilon)}{(\omega-\ci\lambda_L/2-\varepsilon-\Sigma(\omega-\ci\lambda_L/2))(\omega+\ci\lambda_L/2-\varepsilon-\Sigma(\omega+\ci\lambda_L/2))}\notag\\
\Kzinter(\omega)=&\int_{-\Lambda}^\Lambda\de\frac{g(\varepsilon)}{(\omega-\ci\lambda_L/2-\varepsilon-\Sigma(\omega-\ci\lambda_L/2))(\omega+\ci\lambda_L/2+\varepsilon-\Sigma(\omega+\ci\lambda_L/2))},
\end{align}
where we have substituted the expression for $G^{A(R)}_{\pm}(k,\omega\pm\ci\lambda_L/2)$ by analytically continuing eqn.~(7) (main text). The self-energy $\Sigma(\omega\pm\ci\lambda_L/2)$ can be obtained from the spectral function $\rho_\Sigma(\omega)=(-1/\pi)\mathrm{Im}[\Sigma(\omega^+)]$ by using the integral representation
\begin{align}
\Sigma(z=\omega\pm\ci\lambda_L/2)=\int_{-\infty}^{+\infty}\dw\frac{\rho_\Sigma(\omega)}{z-\omega}.
\end{align}
The above equations for $\Kzintra$ and $\Kzintra$ also tells us that the information from the lattice, like dimension etc., enters implicitly through the density of states $g(\varepsilon)$, hence the obtained expressions are general and should be valid for arbitrary dimensions $d\geq 1$ as well.
\subsection{Numerical details}\mylabel{subsec:sup:lyup_numdetails}
\ifdefined\showfigures
\begin{figure}
\includegraphics[scale=1.25]{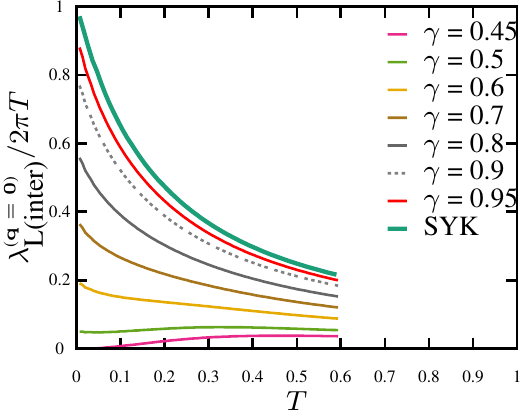}
\caption{{\bf Interband Lyapunov exponent:} Plot of $\lzinter/2\pi T$ as a function of temperature $T$ for $q=2$ and $\gamma=0.45 - 0.95$. The curve for the 0-dim SYK is shown using a bold line.  }
\mylabel{fig:sup:figure7}
\end{figure}
\fi 
A solution for the Lyapunov exponent can now be found by searching for a $\lambda_L$ between $[0,2\pi/T]$ such that the eigenvalue equations (\eqn{eq:sup:OTOCevaleq1} and \eqn{eq:sup:OTOCevaleq2}) get satisfied. To this end we discretize the interval $[-\Omega,\Omega]$, where $\Omega>0$ is a cutoff, into $N_\omega$ divisions and diagonalize the resulting $2N_\omega\times2N_\omega$ matrices, given by
\begin{align}
\MKzintra(\lambda_L)_{i,j}=&\Delta\omega\frac{J^{2}}{2\pi}\Kintra(q,\omega_i)\left[
\begin{array}{cc}
2g_{1}(\omega_i-\omega_j) & -g_{2}(\omega_i-\omega_j)\\
-g_{2}(\omega_i-\omega_j) & 2g_{1}(\omega_i-\omega_j)
\end{array}
\right]\notag\\
\MKzinter(\lambda_L)_{i,j}=&
\Delta\omega
\frac{J^{2}}{2\pi}
\Kinter(q,\omega_i)\left[
\begin{array}{cc}
2g_{1}(\omega_i-\omega_j) & -g_{2}(\omega_i-\omega_j)\\
-g_{2}(\omega_i-\omega_j) & 2g_{1}(\omega_i-\omega_j)
\end{array}
\right],
\end{align}
where $\Delta\omega=2\Omega/N_\omega$ and $\omega_i=-\Omega+(i-1/2)\Delta\omega$ with $i,j$ going from $0,\cdots,N_\omega-1$. We diagonalize the above matrices individually and look for $\lambda_L$s that produce an eigenvalue $1$.  Since we have two eigenvalue equations for the inter/intra band modes we get two solutions for the Lyapunov exponent as well, which we denote as $\lzintra$ and $\lzinter$. As mentioned in the main text we find from numerics that the larger of the two exponents, i.e.
\begin{align}
\lmax=\max\left\{\lzintra, \lzinter\right\}
\end{align}
turns out to be the \emph{intraband} one. We have already shown the behavior of $\lmax$ as a function of temperature for various $\gamma$ in \fig{fig:figure3} of the main text and we do the same for $\lzinter$ in \fig{fig:sup:figure7}. We find that as $\gamma\rightarrow 1$, $\lzinter$ like $\lmax$ starts to merge with the SYK curve. Infact as $\gamma\rightarrow 1$ $\lzinter$ and $\lzintra$ become approximately equal. Also unlike the intraband mode for which a solution $\lzintra$ of \eqn{eq:sup:OTOCevaleq1} always exists for all $\gamma\in(0,1)$, we find that the eigenvalue equation for the interband mode (\eqn{eq:sup:OTOCevaleq2}) ceases to have a solution (within numerical accuracy) when $\gamma$ is approximately less than $0.45$.

\fi 

\end{document}